\documentclass[prd,aps,nofootinbib]{revtex4}
\usepackage{amssymb,amsmath,epsfig}
\usepackage[bookmarks]{hyperref}
\usepackage[usenames,dvipsnames]{color}

\usepackage{graphicx}
\usepackage{ifpdf}
\ifpdf
	\DeclareGraphicsExtensions{pdf,png,jpg}
\else
	\DeclareGraphicsExtensions{.eps}
\fi

\begin{document}

\title{On the behavior of causal geodesics on a Kerr-de Sitter spacetime}

\author{J. Felix Salazar and Thomas Zannias}

\affiliation{ Instituto de F\'\i sica y Matem\'aticas,
Universidad Michoacana de San Nicol\'as de Hidalgo,\\
Edificio C-3, Ciudad Universitaria, 58040 Morelia, Michoac\'an, M\'exico.} 
\email{ jfelixsalazar@ifm.umich.mx, zannias@ifm.umich.mx}

\begin{abstract}
We analyze the behavior of causal geodesics on a Kerr-de Sitter spacetime
with particular emphasis 
on their 
completeness property.
We set up an initial value problem (IVP)
whose solutions lead  to a global understanding 
of causal geodesics on these spacetime.
Causal geodesics that avoid the rotation axis are complete except the ones that
hit the ring-like curvature singularity and those
that encounter
the ring singularity are necessary equatorial ones.
We also  show the existence of geodesics that
cross or lie on the rotation axis.
The equations governing
 the latter family show  the 
repulsive nature of the 
ring singularity.
The results of this work show, that as far as properties of causal geodesics are
concerned, Kerr-de Sitter spacetimes 
behave in a similar manner as
the family of Kerr spacetimes.
\end{abstract}

\date{\today}


\maketitle
\section{Introduction}
 Recent observations of type Ia supernovae \cite{Riess}, \cite{Perlm} suggest that 
 we may live in a universe possessing 
 a  positive cosmological constant $\Lambda$. Even though the data constrain $\Lambda$ in the tiny range  $0<\Lambda <10^{-55} cm^{-2}$
 and thus its effects on the  galactic and sub-galactic structures are negligible,
 nevertheless a non vanishing $\Lambda$ influences the large scale structure of the universe.
  The conformal boundary $\mathcal{J}$ of any spacetime  $(M,g)$ solution
 of Einstein's equations in the presence of $\Lambda$ 
  is spacelike, timelike or null  depending whether $\Lambda$ is positive, negative or zero
  and this property of $\mathcal{J}$
  influences  the asymptotic structure of a spacetime modeling 
  an isolated gravitating system. As a consequence, properties of 
  gravitational radiation, the definition of Bondi four momentum, asymptotic symmetries etc  
 have to be made compatible with the spacelike or timelike character of $\mathcal{J}$. For a recent discussion on these problems 
consult  \cite{Ast} and further references within.\\ 
A non vanishing $\Lambda$ 
can also affect the end state of  the complete gravitational collapse of a bounded system
and thus 
 issues regarding  the cosmic censorship conjecture, the precise definition of 
a black hole 
 have also to be re-accessed (for recent work on these problems see for instance   \cite{Shap}, \cite{Lake}). 

Exact solutions of the Einsteins equations in the presence of a $\Lambda$-term
offer valuable insights into the dynamics of the theory.
Fortunately a non trivial family of such  
solutions has been discovered long ago by Carter
 \cite{Car1},\cite{Car2}. It is a three parameters family 
 characterized by 
 $(\Lambda, M, a)$ where $M$ is interpreted as
 the mass of the solution and $a$ as a rotation parameter (
 for recent advances supporting these interpretations 
of $(M, a)$, consult refs. \cite{Tod}, \cite{Chr2})). For  $\Lambda >0$ the
solutions
 are referred as the Kerr-de Sitter, for $\Lambda<0$ 
as Kerr-anti de-Sitter,  for $\Lambda=0$, they reduce to the Kerr family
 while for $(\Lambda\neq0, a=0)$, they describe the  Schwarzschild-(anti) de Sitter
 family.\\  
 
 For particular values of $(\Lambda, M, a)$,
   Gibbons and Hawking
 \cite{GibHaw1}  interpreted the Kerr-de Sitter solution as describing
 a  black hole in an asymptotically de Sitter background. The structure of 
this type of black holes is different than the structure of the asymptotically 
flat family of Kerr black holes.
The former may  possess up to four horizons, two of them
 are cosmological while the other two are the inner and outer black hole horizons
 enclosing a ring-like singularity (see for instance discussion in  \cite{GibHaw1},\cite{Mat1}).
Clearly the structure
 of these black holes is
 distinct to those exhibited by a Kerr black hole. \\ 
 
In a recent work \cite{LakZan}, it was shown that depending upon 
 the values of the parameters\footnote{ The analysis in \cite{LakZan},
 assumed a tiny value of $\Lambda$ 
 and ranges for $M$ and $a$ describing astrophysical sources. However,
 the classification discussed in \cite{LakZan} is actually independent of these  restrictions.} $(\Lambda, M, a)$,
 the Kerr-de Sitter family of spacetimes $(M,g)$ can describe :\\

a) a black  hole embedded 
within two cosmological horizons,\\ 
 
b) an extreme black hole that find itself within a pair of cosmological horizons,\\ 
  
c) a configuration
where the inner, the outer and one of the 
 cosmological horizons coincide,\\  

d) a spacetime where the outer black hole horizon coincides with one of the cosmological horizons.\\

e) a ring like curvature singularity enclosed within two  cosmological horizons\footnote{
In this classification, the term horizon should be understood as a Killing horizon.}.\\

These result show that the family of Kerr-de Sitter
spacetimes  bears a close resemblance
 to the Reissner-Nordstrom-de Sitter family.
 The Reissner-Nordstrom-de Sitter family, is a three parameter family 
 of spherically symmetric solutions of the Einstein-Maxwell-$\Lambda$ system 
 characterized  by the mass parameter $M$, the electric charge $Q$  and a non vanishing
 $\Lambda$.
 Depending upon the values of 
 $(M,Q,\Lambda)$ the Einstein-Maxwell-de Sitter
 family  describe spacetimes whose horizon structure is identical to 
 the spacetimes in the categories $a)-e)$ above 
  and the  reader is refereed to \cite{Bril}
for further discussion of the Einstein-Maxwell-de Sitter family.\\

Although the Kerr-de Sitter family of spacetimes has been the subject of 
 many investigations,  
these investigations were restricted to a Kerr-de Sitter spacetimes
 that describe a black hole within a pair of cosmological horizons, i.e.
spacetimes belonging to the category a). Geodesics
motion has been the focus of many of these investigations.
For instance equatorial geodesics
have been discussed in \cite{Stuc1} \cite{Stuc2} \cite{Stuc3}
while null geodesics 
studied in  \cite{Kra1},\cite{Kra2},\cite{Kra3}.
In ref.\cite{Laz}, 
solutions of the geodesic equations in Kerr-de Sitter and Kerr-anti-de Sitter spacetimes 
in terms of Weierstrass elliptic and Kleinian functions were presented. The work in \cite{Laz}
 has some intersection with the present work and further ahead we shall comment
 on  the two approaches.\\

In this paper we study causal geodesics on a Kerr-de Sitter spacetime
belonging
to one of the categories $a)-e)$ listed above.
We introduce a 
 Boyer-Lindquist chart
 which even though covers a limited
region of any  Kerr-de Sitter spacetime, nevertheless
the relatively simple form of the metric permit us to gain 
 insights on the  behavior of causal geodesics.
In this work, the region covered by
a Boyer-Lindquist chart will be referred as a Carter's block\footnote{In the analysis of the Kerr metric,
a Carter's block is nothing more 
than a Boyer-Lindquist block (this terminology used widely in ref. \cite{Neil}). Since in this work the background spacetime is a Kerr-de Sitter
and in order to avoid confusion,
we refer to these blocks as Carter's blocks.} and this region is
contained within Killing horizons or a Killing horizon and the asymptotic regions.
Relative to Boyer-Lindquist coordinates, the Hamilton-Jacobi equation 
for geodesic motion  
is completely separable and through the first integrals
we set up an initial value problem (IVP)
whose solutions describe geodesics
within a Carter«s block.
Via an analysis
of the solutions of this IVP,  we show that any causal geodesics with an initial
point off the rotation axis or the ring singularity, exhibits one of the following behaviors:\\

1)  the  geodesics is complete and remains within 
a single Carter«s block\\

2)  the geodesics  within a finite amount of affine parameter 
reaches a
Killing horizon and subsequently is continued as a geodesic  through the Killing horizon
into an adjacent 
Carter«s block\\
 
3) the geodesics only asymptotically (ie only after infinite amount of affine parameter)
 reaches a
Killing horizon\\
 
4) the geodesic within a finite amount of the  affine parameter runs into the ring singularity.\\

Moreover
we show that the only causal geodesics that
 reach the ring  singularity are necessarily equatorial ones
 and these geodesics require 
only a finite amount of affine time
to reach the ring singularity.
We also examine the properties of geodesics
that either cross or lie entirely
on the rotation axis. For this analysis, we introduce a local coordinate system 
that covers the rotation axis and show that there exist geodesics that 
cross the rotation axis.
Although we believe that these geodesics are complete,
 we have been unable to provided a  proof of this property.
Geodesics that lie entirely on the axis satisfy a relatively simple set of equations
and through an analysis of these equations we present evidences regarding the repulsive nature of the 
ring singularity.
We find that no geodesics
can pass through the ring singularity 
 unless their ``energy parameter`  is greater than some minimum value 
 which is determined by the height of a repulsive 
 effective potential. The structure of this effective potential  differs from the potential associated to
 the familiar Kerr case.\\

The results of this work show, that
causal geodesics on a Kerr-de Sitter spacetime behave
in a manner analogous to those 
for the case of Kerr (see the fundamental paper by Carter ref. \cite{Car3} also ref. \cite{Neil}).
For instance, while for  the Kerr background, the function
$R(r)=P^{2}(r)-{\Delta(r)}(m^{2}r^{2}+l^{2})$, with $P(r)=(r^{2}+a^{2})E-al_{z}$ and ${\Delta(r)}=r^{2}-2Mr +a^{2}$
plays an important role in the description of geodesics, we find that for the case
of Kerr-de Sitter it is the function
$R(r)=I^{2}P^{2}(r)-{\Delta(r)}(m^{2}r^{2}+l^{2})$
where $\Delta(r) :=-\frac {1}{3}\Lambda r^2(r^{2}+a^{2})+r^{2}- 2M r + a^2$
and $I:=1+\frac {1}{3}\Lambda a^2$
that play an analogous role.
For the case of $\Lambda>0$, the regions where $R(r)$
is positive exhibits more diversity in comparison to the case
of a Kerr background, but nevertheless
the qualitative behavior of causal geodesics for both backgrounds
are those outlined in $1)-4)$ above.
On the other hand, since for both backgrounds the real roots of  $\Delta(r)=0$
determine
 the location of Killing horizons, 
 clearly a non vanishing $\Lambda$ can change drastically the structure of the spacetime
 as the classification $a)-e)$ shows.\\
 
 The plan of the paper is as follows:
In the next section, a brief introduction to the family of Kerr-(anti) de Sitter spacetimes is presented
and in section $(II)$ the first integrals of geodesic motion on Kerr- (anti) de Sitter  spacetimes 
are derived. The derivation begins with the separability
of  the Hamilton-Jacobi equation in a  local Boyer-Lindquist coordinates.
Based on these first integrals, in section $IV$, we define  a suitable initial value problem (IVP)
whose solutions shed light on the behavior of causal geodesics.
In section  $(V)$, we construct
the principal null vector fields and based on these fields we introduce generalized Finkelstein coordinates. 
Relative to these coordinates the metric is free of coordinate singularities
 and using these coordinates we study the continuation
 of causal geodesics through Killing horizons.
Finally the last section of the paper deals with geodesics crossing or lying entirely on the rotation axis
and in the conclusion section a brief discussion of future work
is outlined. In the two  Appendixes some technicalities regarding the IVP problem of section $IV$
are addressed. 

\section{The Kerr-(anti) de Sitter metric }
In this section and for orientation purposes, we provide a brief introduction 
to  the Kerr-(anti) de Sitter family of metrics
(for additional discussion and references, the reader is referred to refs. \cite{GibHaw1},\cite{Mat1},\cite{Chr1},\cite{Olz}).
In local Boyer-Lindquist coordinates, this family is described by: 

\begin{equation}
g = -\frac {\Delta (r)}{I^{2}\rho^{2}}[dt-asin^{2}{\theta}d\varphi]^{2}+
\frac {\hat {\Delta}(\vartheta)\sin^{2}\vartheta}{I^{2}\rho^{2}} [ a dt - (r^{2}+a^{2})d\varphi]^{2} 
+\frac {\rho^{2}} {\Delta(r)}dr^{2}+ \frac {\rho^{2}} {\hat {\Delta}(\vartheta)}d\vartheta^{2}
\label{Eq:g}
\end{equation}
$$
\rho^2 := r^2 + a^2\cos^2\vartheta,\qquad
\Delta(r) :=-\frac {1}{3}\Lambda r^2(r^{2}+a^{2})+r^{2}- 2M r + a^2,\qquad {\hat \Delta}(\vartheta) :=1+\frac {1}{3}\Lambda a^2cos^{2}\vartheta,\qquad I:=1+\frac {1}{3}\Lambda a^2
$$
where $\Lambda>0$ corresponds to Kerr-de Sitter metric, $\Lambda< 0$
the Kerr-anti de Sitter family and $\Lambda=0$  the Kerr metric. 
The $t$-coordinate takes its  values over the real line, 
the angular coordinates $(\vartheta, \varphi)$ vary in the familiar range, while 
$r$  is restricted to a suitable open set of the real line to be made precise further bellow.
The fields
$\xi_{t}=\frac {\partial }{\partial t}$ and  $\xi_{\varphi}=\frac {\partial }{\partial \varphi}$
are commuting Killing  fields
with the zeros of $g(\xi_{t}, \xi_{t})$ determining the location of ergospheres
while the zeros 
of $\xi_{\varphi}$ the rotation axis.
Algebraic manipulations via $GRTensorII$ \cite{grt}, 
show that
the two independent Weyl scalar invariants for (\ref{Eq:g}) have the form:
\begin{equation}
C_{\mu\nu\lambda\rho}C^{\mu\nu\lambda\rho}=\frac{48M^{2}}{\rho^{12}}F(r,\vartheta), \quad F(r,\vartheta)=(r^{2}-a^{2}cos^{2}\vartheta)(\rho^{4}-16a^2r^{2}cos^{2}\vartheta)
\label{Eq:WS}
\end{equation}
\begin{equation}
C^{*}_{\mu\nu\lambda\rho}C^{\mu\nu\lambda\rho}=\frac{96M^{2}ra}{\rho^{12}}F^{*}(r,\vartheta), \quad F^{*}(r,\vartheta)=(r^{2}-3a^{2}cos^{2}\vartheta)(-3r^{2}+a^{2}cos^{2}\vartheta)cos\vartheta
\label{Eq:WDS}
\end{equation}
where 
$C^{*}_{\mu\nu\lambda\rho}$ are the components of the dual of the Weyl tensor.
These invariants, show that the curvature of  (\ref{Eq:g})
becomes unbounded as $\rho \to 0 $ i.e. as the ring  ($r= 0$, $\vartheta= \frac {\pi}{2}$)
is approached. \\

Coordinate singularities in
the components of $g$  occur along 
the rotation
axis ie at  $sin\vartheta=0$ and these  singularities\footnote{Notice that  for  $\Lambda<0$ and for particular values
of the rotation parameter $a^{2}$,
a peculiar coordinate singularity arises at the zeros of ${\hat \Delta}(\vartheta)=0$ or (and)
at the zeros of $I=1+\frac {1}{3}\Lambda a^2=0$. However for $\Lambda>0$
these singularities are absent.}
 also can be removed by introducing generalized Kerr-Schild coordinates
or suitable local coordinates (see section $VI$).
Singularities in the components of (\ref{Eq:g}) also occur at the roots of the quartic equation $\Delta(r)=0$ 
and these are also coordinate singularities
marking the location  of Killing horizons. Depending upon the parameters $(M,\Lambda, a)$ the quartic equation $\Delta(r)=0$ may admit up to four distinct real roots and the well studied Kerr-de Sitter spacetime 
corresponds to a particular choice of $(M,\Lambda, a)$
so that  $\Delta(r)=0$ admits four distinct real roots.
 However, as discussed in \cite{LakZan}, other choices of the parameters $(M,\Lambda, a)$ 
 lead to less than four real roots or to situations where
 some roots are double or exhibit higher multiplicity.
 This in turn leads to  Kerr-de-Sitter  spacetime admitting
 one or more degenerate  Killing horizons.
 Extendability of  the metric 
 across these singularities will be discussed 
in section $V.$\\

In the next section, we derive the first integrals of geodesic motion
 relative to  Boyer-Lindquist coordinates. 
 Our choice of the 
  Boyer-Lindquist coordinates
  has been motivated by the relatively simple form of  the metric $g$ 
  which in turn leads  to simple forms
 for the  first integrals for geodesic motion.
 The draw back of these coordinates
  lies in the 
 property that geodesics reach the Killing horizons and thus are required to be continued
 through these horizons a process discussed in section $V.$\\

 For the continuation of geodesics  through  Killing horizons,
it is convenient to introduce 
 the notion of a Carter's block\footnote{see footnote $3$.}
A Carter's block, denoted here after as $(T,g)$, is a spacetime
covered by a single Boyer-Lindquist chart $(t,r,\vartheta,\varphi)$
so that the metric $g$  is described by (\ref{Eq:g})
but now the  $r$-coordinate take its values in the open interval $(r_{i},r_{i+1})$ 
where $r_{i}$ and  $r_{i+1}$ 
are  consecutive roots\footnote{
The emphasis in the present work 
is devoted to the case where $\Lambda>0$
even though in many occasions we 
quote results valid for the case 
$\Lambda<0$. Without further notice
consecutive real roots of $\Delta (r)=0$ are
denoted by  $r_{i}$,  $r_{i+1}$
and  for the case where $\Delta (r)=0$ admits four distinct real roots
these roots are taken in the form:
$r_{1}<0<r_{2}<r_{3}<r_{4}.$} of $\Delta (r)=0$. The 
 maximal extension of
a Kerr-(anti) de Sitter spacetime is obtained  by assembling 
Carters block in the same manner 
as we obtain the maximal extension of
the Kerr manifold by assembling 
distinct Boyer-Lindquist blocks (see for instance \cite{Car3}, \cite{Neil}). 

Notice however that  in general,
more than three blocks are required to build the maximal extension
of a Kerr-(anti) de Sitter spacetime.
For instance, if  the parameters  $(\Lambda,M,a)$ are chosen so that
$\Delta (r)=0$ admits four real roots $r_{1}<0<r_{2}<r_{3}<r_{4}$
 then  any $(T, g)$ block can be specified
 by restricting the values of the  $r$ 
  in one of the intervals: $(-\infty, r_{1}), (r_{1},r_{2}), (r_{2},r_{3}), (r_{3},r_{4}), (r_{4},\infty)$.
  In this case five building blocks are needed for the construction of  the maximally
 extended  spacetime. The block $(T, g)$ where $r \in  (r_{4},\infty)$ or 
$r \in  (-\infty, r_{1})$, define the two asymptotic blocks while
the block $(T, g)$ with $r \in  (r_{1},  r_{2})$ is the block that contains  the ring singularity.
When $\Delta (r)=0$ admits multiple roots (or roots in the complex plane),
 the number of Carters blocks required for the construction of the maximal extension are
reduced as well.

\section{Separability of the Hamilton-Jacobi equation }
In  this  section, we consider an arbitrary block $(T,g)$ and show that the Hamilton-Jacobi equation 
for geodesics motion is completely separable. 
 Although this separability property has been discussed in the literature, for completeness purposes, we provide 
 a brief discussion leading to the construction of the first integrals of geodesic motion.

For an arbitrary block $(T,g)$,
we write the metric $g$ in (\ref{Eq:g}) in the form: 
\begin{equation}
g = g_{tt}dt^{2}+2g_{t\varphi}dt d{\varphi}+g_{\varphi\varphi}d{\varphi}^{2}+g_{rr}dr^{2}+g_{\vartheta\vartheta}d{\vartheta}^{2}.
\label{Eq:gp}
\end{equation}
where:
\begin{equation}
 g_{tt}=-\frac {{\Delta({r})}-{{\hat \Delta}}({\vartheta})a^{2}sin^{2}\vartheta}{I^{2}\rho^{2}},\quad g_{\varphi \varphi}=
 \frac {{\hat \Delta}({\vartheta})(r^{2}+a^{2})^{2}-\Delta(r)a^{2}sin^{2}\vartheta}{I^{2}\rho^{2}}sin^{2}\vartheta,
\quad g_{t\varphi}=\frac {\Delta(r)-{\hat \Delta}({\vartheta})(r^{2}+a^{2})}
{I^{2}\rho^{2}}asin^{2}\vartheta
\end{equation}
\begin{equation}
g_{rr}=\frac {\rho^{2}}{\Delta(r)},\quad\quad g_{\vartheta \vartheta}=\frac {\rho^{2}}{\hat \Delta(\vartheta)}
\end{equation}
The non vanishing components $g^{\mu\nu}$ of the inverse metric $g^{-1}$
are:
\begin{equation}
g^{tt}=\frac {g_{\varphi\varphi}}{det \hat g}=-\frac {I^{2}[{\hat {\Delta}(\vartheta)}(r^{2}+a^{2})^{2}-{\Delta(r)}a^{2}sin^{2}\vartheta]}{\rho^{2}{\hat {\Delta}(\vartheta)}{\Delta(r)}},\quad
g^{t\varphi}=-\frac {g_{t\varphi}}{det \hat g}=\frac {I^{2}a[{\Delta(r)}-{\hat {\Delta}(\vartheta)}(r^{2}+a^{2})]}{
\rho^{2}{\hat {\Delta}(\vartheta)}{\Delta(r)}}
\label{Eq:af}
\end{equation}
\begin{equation}
g^{\varphi\varphi}=\frac {g_{tt}}{det \hat g}=\frac {I^{2}[{\Delta(r)}-{\hat {\Delta}(\vartheta)}a^{2}sin^{2}\vartheta]}{
\rho^{2}sin^{2}\vartheta{\hat{\Delta}(\vartheta)}{\Delta(r)}},\quad g^{rr}=\frac {\Delta{(r)}}{\rho^{2}},
 \quad  g^{\vartheta\vartheta}=\frac {\hat {\Delta} {(\vartheta)}}{\rho^{2}} 
 \label{Eq:af}
\end{equation}
where 

$$det \hat g=g_{tt}g_{\varphi\varphi}-(g_{t\varphi})^{2}=-\frac {sin^{2}\theta{\Delta(r)}{\hat {\Delta}(\theta)}}{I^{4}}.$$

The equations of geodesics
motion can be derived from the Hamiltonian function\footnote{For a more thorough discussion of this Hamiltonian 
and the associated Hamilton-Jacobi equation within the framework of a symplectic phase space see for instance (\cite{O1},\cite{O2}, \cite{O3})}
\begin{equation}
H(x^{\mu}, p_{\mu})=\frac {1}{2} g^{\mu\nu}p_{\mu}p_{\nu}
\label{Eq:Hf}
\end{equation}
subject to the constraint $H(x^{\mu}, p_{\mu})=-m^{2}$ with $m>0$ the rest mass of the
test particle while $m=0$ describes zero zero rest mass particles. 
Accordingly, the Hamilton-Jacobi equation has the form 
\begin{equation}
 g^{\mu\nu}\frac {\partial S}
 {{\partial x^{\mu}}}\frac {\partial S}
 {{\partial x^{\nu}}} =-m^{2},\quad p_{\mu}=\frac {\partial S}{{\partial x^{\mu}}} 
\label{Eq:HJ}
\end{equation}
and the symmetries of the background metric,
suggest the ansatz
\begin{equation}
 S(x^{\mu}, p_{\mu})= -Et+l_{z}\varphi+\hat S(r,\vartheta)
 \label{Eq:HF}
\end{equation}
where $(E, l_{z})$ are constants. Substituting this ansatz into 
(\ref{Eq:HJ}), we find after some algebra:
\begin{equation}
\frac {I^{2}}{\rho^{2}}\left[-\frac {(r^{2}+a^{2})^{2}}{{\Delta(r)}}+\frac {a^{2}sin^{2}{\vartheta}}{{\hat {\Delta}(\vartheta)}}
\right]E^{2}-2\frac {I^{2}a}{\rho^{2}}\left[\frac {1}{{\hat {\Delta}(\vartheta)}}-\frac {r^{2}+a^{2}}{{\Delta(r)}}
\right]El_{z}+\frac {I^{2}}{\rho^{2}}
\left[\frac {1}{{\hat {\Delta}(\vartheta)}sin^{2}\vartheta}-\frac {a^{2}}{{\Delta(r)}}
\right]l^{2}{}_{z}+\nonumber
\end{equation}
\begin{equation}
+\frac {{\Delta(r)}}{{\rho^{2}}}(\frac {{\partial {\hat S}}}{{\partial r}})^{2}+\frac {{\hat {\Delta}(\vartheta)}}{{\rho^{2}}}(\frac {\partial {\hat S}}
{\partial \vartheta})^{2}=-m^{2}
\label{Eq:IHJ}
\end{equation}
Multiplying this equation by $\rho^{2}$, setting ${\hat S}(r, \vartheta)={\hat S_{r}}(r)+
{\hat S_{\vartheta}}(\vartheta)$ and remembering that $\rho^{2}=r^{2}+a^{2}cos^{2}\vartheta$ we get:
\begin{equation}
 {\Delta(r)} (\frac {{\partial {\hat S_{r}}}}{{\partial r}})^{2}-\frac {I^{2}}{ {\Delta(r)}} \left[(r^{2}+a^{2})E-al_{z}\right]^{2}+m^{2}r^{2}= 
 -{{\hat \Delta}(\vartheta)} (\frac {{\partial {\hat S_{\vartheta}}}}{{\partial \vartheta}})^{2} -\frac {I^{2}}{{\hat {\Delta}(\vartheta)}}
 \left[\frac {l_{z}}{sin^{2}{\vartheta}}-asin{\vartheta}E\right]^{2}-m^{2}a^{2}cos^{2}{\vartheta} 
  \label{Eq:HF}
\end{equation}
Since $I$ is a constant, after rearrangement we obtain
\begin{equation}
  {\Delta(r)}^{2} (\frac {{\partial {\hat S_{r}}}}{{\partial r}})^{2}-{I^{2}} \left[(r^{2}+a^{2})E-al_{z}\right]^{2}+{\Delta(r)} [m^{2}r^{2}+l^{2}]=0  
  \label{Eq:HF}
\end{equation}
\begin{equation}
-{\hat {\Delta}(\vartheta)} (\frac {{\partial {\hat S_{\vartheta}}}}{{\partial \vartheta}})^{2} -\frac {I^{2}}{{{\hat \Delta}(\vartheta)}}
 \left[\frac {l_{z}}{sin^{2}{\vartheta}}-asin{\vartheta}E\right]^{2}-m^{2}a^{2}cos^{2}{\vartheta}+l^{2}=0 
  \label{Eq:HF}
\end{equation}
 where the separation constant  
$l^{2}$ is the  Carter«s constant (denoted by $\mathcal{K}$ in Carters original paper \cite{Car3}). 
We introduce the functions $\Theta(\vartheta)$, $R(r)$ and write these equations in the form:  
 \begin{equation}
 (\frac {{\partial {\hat S_{r}}}}{{\partial r}})^{2}=\frac {R(r)}{ {\Delta(r)}^{2}},\quad 
 (\frac {{\partial {\hat S_{\vartheta}}}}{{\partial \vartheta}})^{2}=\frac {\Theta(\vartheta)}{{\hat {\Delta}(\vartheta)}^{2}} 
  \label{Eq:HF}
\end{equation}
 \begin{equation}
R(r)=I^{2} \left[(r^{2}+a^{2})E-al_{z}\right]^{2}-{\Delta(r)}(m^{2}r^{2}+l^{2})
   \label{Eq:R}
\end{equation}
 \begin{equation}
\Theta(\vartheta)= {\hat {\Delta}(\vartheta)}l^{2}- I^{2}\left[\frac {l_{z}}{sin{\vartheta}}-asin{\vartheta}E\right]^{2}-{\hat \Delta}(\vartheta)m^{2}a^{2}cos^{2}{\vartheta}.
    \label{Eq:THH}
\end{equation}
 The function
$\Theta(\vartheta)$ in the limit that $\Lambda\to 0$
agrees with eq $(53)$  in Carters
paper provided  $l_{z}$ is replaced by  $\Phi$ 
and moreover introduce a new constant $Q$ by $Q=l^{2}-I^{2}(l_{z}-aE)^{2}$.
Moreover at the same limit,
the function  $R(r)$ reduces to eq $(55)$ in Carters paper\footnote{Altough, Carter in \cite{Car3} separated the Hamilton-Jacobi
equation on a Kerr-Newmann background relative to a Finkelstein  coordinate system 
nevertheless the functions $\Theta(\vartheta)$ and $R(r)$ appearing in eqs
$(53-55)$ in his paper, in the limit of vanishing $\Lambda$ and electric charge,
agree with equations 
 (\ref{Eq:R}, \ref{Eq:THH}).}.\\  
The separability of the Hamilton-Jacobi implies that
timelike or null geodesics on the background of (\ref{Eq:g})
are described by:
\begin{equation}
\rho^{2} \frac {dt}{d\lambda}=\frac {I^{2}[{\hat \Delta}(\vartheta)(r^{2}+a^{2})^{2}-{\Delta(r)}a^{2}sin^{2}\vartheta]}
{{\hat \Delta}(\vartheta){\Delta(r)}}E
 +\frac {I^{2}a[{\Delta(r)}-{{\hat \Delta}(\vartheta)}(r^{2}+a^{2})]}{
{{\hat \Delta}(\vartheta)}{\Delta(r)}}l_{z} 
  \label{Eq:GT}
\end{equation}

\begin{equation}
\rho^{2} \frac {d\varphi}{d\lambda}=
\frac {I^{2}[{\Delta(r)}-{\hat \Delta}(\vartheta)a^{2}sin^{2}\vartheta]}
{sin^{2}\vartheta{\hat \Delta}(\vartheta){\Delta(r)}}l_{z}-
\frac {I^{2}a[\Delta(r)-{\hat \Delta}(\vartheta)(r^{2}+a^{2})]}
{{\hat \Delta}(\vartheta)\Delta(r)}E
 \label{Eq:GP}
 \end{equation}

\begin{equation}
\rho^{2} \frac {dr}{d\lambda}=\pm {R(r)}^{\frac{1}{2}},\quad \rho^{2} \frac {d\vartheta}{d\lambda}=\pm {\Theta(\vartheta)}^{\frac{1}{2}}
 \label{Eq:YYY}
\end{equation}
 Rearranging the first two equations, we get an equivalent system
\begin{equation}
\rho^{2} \frac {dt}{d\lambda}= \frac {I^{2}(r^{2}+a^2)}{\Delta(r)}[(r^{2}+a^{2})E-al_{z}]+
 \frac {I^{2}a}{\hat {\Delta}(\vartheta)}[l_{z}-aE sin^{2}\vartheta]
 \label{Eq:GTM}
\end{equation}
\begin{equation}
\rho^{2} \frac {d\varphi}{d\lambda}=\frac {I^{2}a}{\Delta(r)}[(r^{2}+a^{2})E-al_{z}] 
 - \frac {I^{2}a}{\hat {\Delta}(\vartheta)}[E-\frac {l_{z}}{asin^{2}\vartheta}]
  \label{Eq:GPM}
\end{equation}
\begin{equation}
\rho^{2} \frac {dr}{d\lambda}=\pm {R(r)}^{\frac{1}{2}}
  \label{Eq:GR}
\end{equation}
 \begin{equation}
\rho^{2} \frac {d\vartheta}{d\lambda}=\pm {\Theta(\vartheta)}^{\frac{1}{2}}
  \label{Eq:GTH}
\end{equation}
which describe the behavior of timelike  ($m>0$) or null ($m=0$) geodesics relative
to an arbitrary block $(T,g)$.\\\

For any initial point\footnote{For the moment, we assume that this initial point is off the rotation axis
and off the ring singularity. For initial points on the axis or on the  ring singularity
 see discussion in section $VI$ and the conclusion section.}
$(t_{0}, \varphi_{0}, r_{0},\vartheta_{0})$ in $(T,g)$
and  set  of constants
 $(l^{2},E,  l_{z}, m \geq 0)$ subject to the restriction that 
 $R(r_{0})>0$ and $\Theta(\vartheta_{0})>0$,
 the system $( \ref{Eq:GTM}-\ref{Eq:GTH})$ defines an initial value problem (IVP)
 and the focus of this work 
 is to analyze solutions of this IVP with particular emphasis
on their completeness property. If $(t(\lambda), r(\lambda),\vartheta(\lambda),\varphi(\lambda))$
is a solution of $( \ref{Eq:GTM}-\ref{Eq:GTH})$
representing a timelike or a null geodesic, then it is 
 a complete geodesic if it can 
 be extended (as a geodesic)  to unbounded values of the affine  parameter $\lambda$.
 However, it is not clear that every solution
 of $( \ref{Eq:GTM}-\ref{Eq:GTH})$
 is a complete geodesic.
 Even if the block under consideration is free of the ring singularity,
 an obstruction to their extendebility 
to all $\lambda \in (-\infty, \infty)$  
 may arise due to the  fact that the geodesic runs into the boundary of $(T,g)$
 (recall for any  $(T,g)$, 
the coordinate $r$ takes its values in intervals of the form $(r_{i},r_{i+1})$).
We shall deal with these problems 
in the next section, but here  we discuss a few properties
 of the functions $R(r)$,
$\Theta(\vartheta)$  that will be useful further ahead.\\

Upon  introducing a new constant  $Q$ via
 \begin{equation}
Q=l^{2}-I^{2}(l_{z}-aE)^{2}
\label{Eq:NC}
\end{equation}
$\Theta(\vartheta)$
can be written as
\begin{equation}
\Theta(\vartheta)= \left[{\hat \Delta(\vartheta)}l^{2}- I^{2}\left[\frac {l_{z}}{sin{\vartheta}}-asin{\vartheta}E\right]^{2}-\hat \Delta(\vartheta)m^{2}a^{2}cos^{2}{\vartheta}\right]=
    \label{Eq:TH}
\end{equation}
\begin{equation}
= Q+I^{2}(a^{2}E^{2}-\frac {l_{z}^{2}}{sin^{2}{\vartheta}})cos^{2}\vartheta-m^{2}a^{2}cos^{2}\vartheta+
\frac {\Lambda a^{2}}{3}(l^{2}-m^{2}a^{2}cos^{2}\vartheta)cos^{2}\vartheta.
\label{Eq:THN}
\end{equation}
Clearly if $\l_{z}\neq0$, then $\Theta(\vartheta)$ diverges to $-\infty$ as 
$\sin\vartheta\to 0$ ie  as the rotation axis is approached. 
Therefore if 
$sin\vartheta_{0}\neq0$,
ie $\vartheta_{0}$ is 
 off the axis and the constants 
 $(l^{2},E,  l_{z}, m \geq 0)$ 
have been chosen so that
$\Theta(\vartheta_{0})>0$, then continuity arguments 
imply that there exist a neighborhood 
$(\vartheta_{1}, \vartheta_{2})$ of $\vartheta_{0}$
such that $\Theta(\vartheta)$ is positive  on $(\vartheta_{1}, \vartheta_{2})$ 
and $\Theta(\vartheta_{1})=\Theta(\vartheta_{2})=0$.
Positivity of $\Theta(\vartheta)$ is a necessary condition for the existence
of solutions of (\ref{Eq:GTH})
and properties of these solutions 
depend upon the value of the constant $Q$.\\

Suppose first that $Q>0$, then from (\ref{Eq:THN}) it follows that  $\Theta(\vartheta=\frac {\pi}{2})=Q$
and as long as $l_{z}\neq 0$,
 any solution of (\ref{Eq:GTH}) oscillate around the equatorial $\vartheta=\frac {\pi}{2} $
plane. If however $l_{z}=0$, and the parameters $(E, m^{2})$
are chosen appropriately, then $\Theta(\vartheta)$ can be made non vanishing on the entire interval $[0,\pi]$.
In this case,
solutions of (\ref{Eq:GTH})
are related to spherical polar and polar geodesics (often referred as orbits), 
and at the end of this section we shall comment
on these type of geodesics (orbits).\\
 
For the choice $Q=0$, one family of solutions 
of (\ref{Eq:GTH}) is 
described by  $\vartheta(\lambda)
=\frac {\pi}{2}$ which leads to the family of equatorial geodesics.
Moreover since the equatorial $\vartheta=\frac {\pi}{2}$ plane,  is a closed, totally geodesic submanifold
consisting of fixed points of the isometry that sends $\vartheta \to \pi-\vartheta$
and leaves the other Boyer-Lindquist coordinates intact (see discussion in \cite{Neil}), 
it follows that if any
solution of  (\ref{Eq:GTH}) has the property that
if $\frac {d\vartheta}{d\lambda}=0$ at $\vartheta=\frac {\pi}{2} $,
 then the entire geodesic  lies on the equatorial plane.
 Therefore non trivial solutions of (\ref{Eq:GTH})
 having $Q=0$ either lie entirely on the equatorial plane 
 or do not intersect it.\\

Finally  for $Q<0$, no solution of  $(\ref{Eq:GTH})$ can cross or touch the equatorial
plane but it can be shown easily that there exist solutions that are confined either to lie ``above`` 
or to lie entirely ``bellow`
the equatorial plane.\\

We now consider the radial function $R(r)$. 
Using the form of $\Delta(r)$ in (\ref{Eq:g}),
we find:
 \begin{equation}
R(r)=I^{2} \left[(r^{2}+a^{2})E-al_{z}\right]^{2}-{\Delta(r)}(m^{2}r^{2}+l^{2})
=\frac {1}{3} \Lambda m^{2}r^{6}+[I^{2}E^{2}+\frac {1}{3} \Lambda l^{2}-(1-\frac {1}{3} \Lambda a^{2})m^{2}
]r^4+
 \end{equation}
\begin{equation}
 +2Mm^{2}r^{3}+[2I^{2}(a^{2}E^{2}-al_{z}E)-[a^{2}m^{2}+l^{2}(1-\frac {1}{3} \Lambda a^{2})]r^{2}+2ml^{2}r+I^{2}a^{2}(aE-l_{z})^{2}-a^{2}l^{2}
\label{Eq:MR}
\end{equation}
which shows for $\Lambda>0$ and $m\neq  0$, asymptotically ie as 
$r\to \pm\infty$, $R(r)$ is positive and thus timelike geodesics may reach  the 
asymptotic region (in contrast to the case where $\Lambda<0$).
For $m=0$, and  $\Lambda>0$, null geodesics can also 
reach the asymptotic
region without imposing any restrictions upon $E$ or upon the Carters constant $l$.
\\ 

Since motion in the radial direction
takes place only in intervals where $R(r)>0$, therefore 
 the location of the real roots of the  equation $R(r)=0$
 are very important. Following Carter,
 we introduce the function $P(r)$ via $P(r)=(r^{2}+a^{2})E-al_{z}$
 so that $R(r)$  can be  written as $R(r)=I^{2}P^{2}(r)-{\Delta(r)}(m^{2}r^{2}+l^{2})$
 and this representation allows us to
 correlate the roots of $R(r)=0$ to those of $\Delta(r)=0$.
\\

Let first $r_{i}$ be a real root
of $\Delta(r)=0$ and let 
one of the constants  $(m,l)$ be different than zero\footnote{For the case where $m=l=0$
the roots of $R(r)=0$ are identical to that of $P(r)=0$ and we shall not examine this case any further.}.
The relation: 
 $R(r)=I^{2}P^{2}(r)-{\Delta(r)}(m^{2}r^{2}+l^{2})$
implies 
 $R(r_{i})=I^{2}P^{2}(r_{i})=
 I^{2} \left[(r_{i}^{2}+a^{2})E-al_{z}\right]^{2}$
and thus $R(r_{i})$ is always positive
except when: $E=l_{z}=0$ or when $P(r_{i})=0$.
Although the constraint $E=l_{z}=0$
seems to be restrictive, actually there exist causal geodesics characterized by  $E=l_{z}=0$
and at the end of the next section, we mention some of their properties.
Excluding for the moment 
the $E=l_{z}=0$ case, it follows
that at any root $r_{i}$ of $\Delta(r)=0$,
 always $R(r_{i})>0$
except for the particular case where $P(r_{i})=0$
and in this event
$R(r_{i})=0$ as well.\\

Let now $\hat r_{1}$ be any real root\footnote{The possibility that all roots of $R(r)$ are complex it is
not a-priori excluded.
In such event, and for $\Lambda>0$, the function $R(r)$ is positive definite over the entire real line} of 
$R(r)=0$ and let again at least one of the $(m,l)$ is different than zero.
From  $R(r)=I^{2}P^{2}(r)-{\Delta(r)}(m^{2}r^{2}+l^{2})$,
we conclude that necessarily
 $\Delta(\hat r_{i})>0$
  which 
means that  any real root of $R(r)=0$ 
occurs in intervals where $\Delta(r)>0.$ 
Whenever $P(r)$ and $\Delta (r)$ share a common root 
denoted by $r_{i}$ then this $r_{i}$ is also a root of $R(r)=0$.
Moreover  if this $r_{i}$ 
is a simple root of $\Delta (r_{i})=0$ then $r_{i}$ is a simple root
  of $R(r)=0$
In sum, whenever $\Delta (r)>0$ on $(r_{i},r_{i+1})$,
then $R(r)=0$ may have zeros on  $[r_{i}, r_{i+1}]$
while in  the case where $\Delta (r)<0$ on $(r_{i},r_{i+1})$ 
then no roots of $R(r)=0$ lie
on $(r_{i}, r_{i+1})$ (although the possibility that both 
$r_{i},r_{i+1}$ are roots of $R(r)=0$ it is not excluded).
Plots of the functions $\Delta(r)$ and $R(r)$
are shown and  various possibilities regarding their roots are 
indicated in Figs.1 and 2.\\

\begin{figure}
\centering
\begin{minipage}[t]{0.5\textwidth}
	\includegraphics[width=0.4\textwidth]{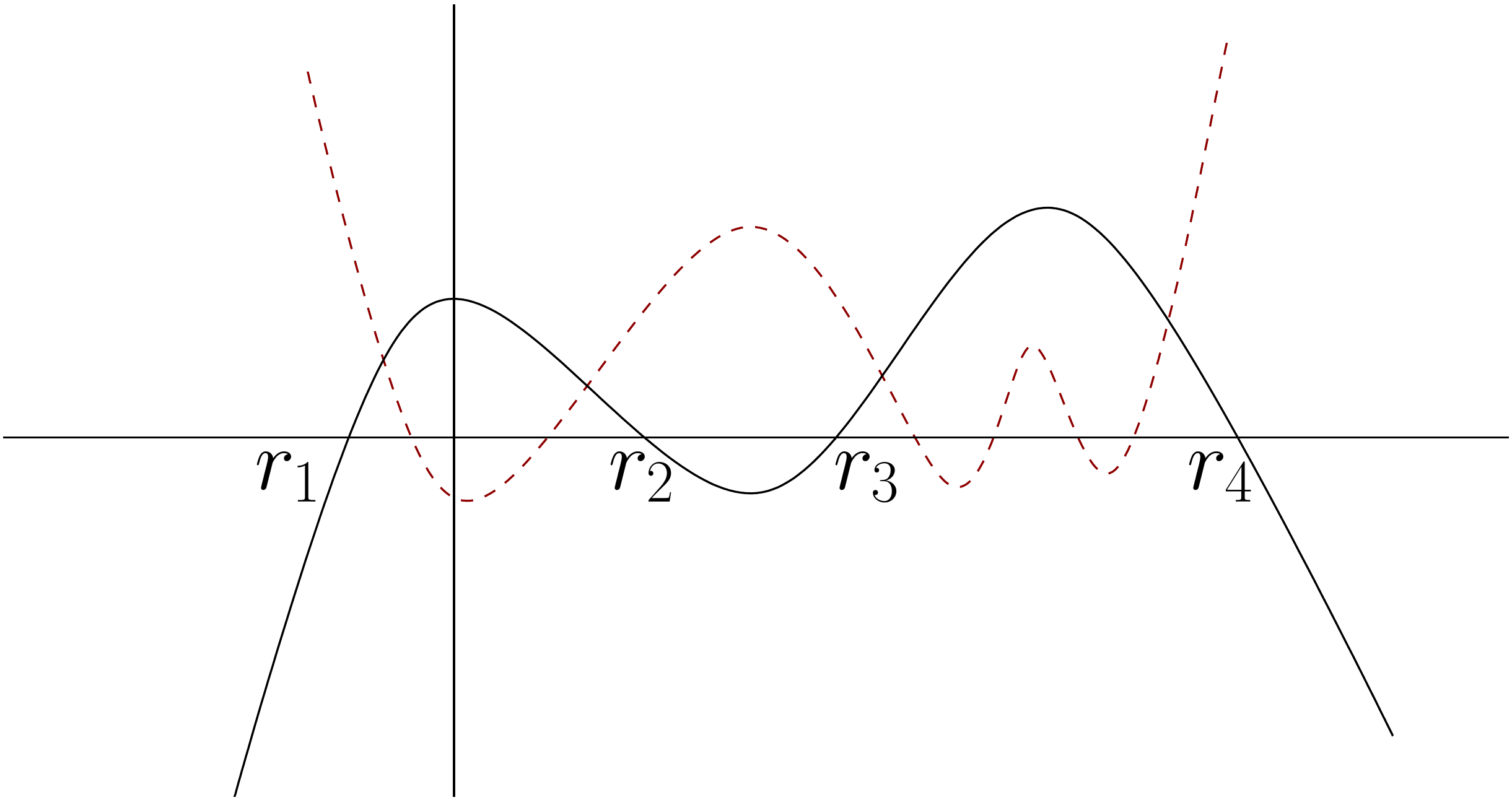}
	\caption{In this figure the blue solid line is the plot of $\Delta(r)$  while the dashed line is the graph of $R(r)$.
	The four distinct roots of $\Delta(r)$ are indicated by $r_{1}<0< r_{2}<r_{3}<r_{4}$. Two roots of $R(r)=0$
	are shown in the block specified by $(r_{1}, r_{2})$ while four other roots are in the block specified by 
	$(r_{3}, r_{4})$. For this plot $\Lambda>0$.}	
\end{minipage}
\qquad
\centering
\begin{minipage}[t]{0.5\textwidth}
	\includegraphics[width=0.5\textwidth]{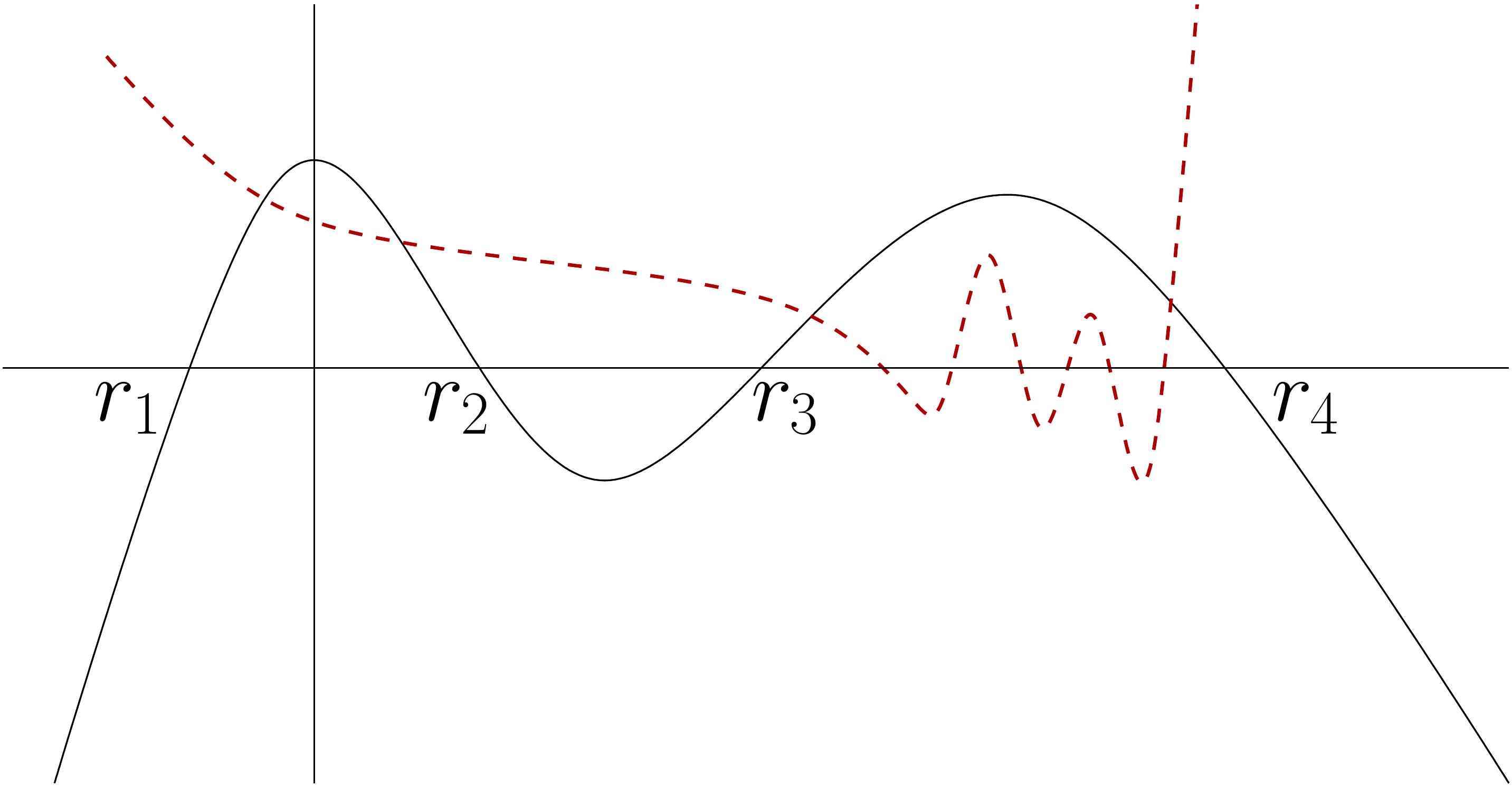}
	\caption{In this figure, as in Fig.1, again $\Lambda >0$, the blue solid line is the plot of $\Delta(r)$  while the        
	dashed line is the graph of $R(r)$.
	Here all real roots of $R(r)=0$ lie in the block specified by 
	$(r_{3}, r_{4}).$}
\end{minipage}
\end{figure}

The properties of the functions $\Theta(\vartheta)$ and $R(r)$ 
 are very useful and here as a first application
 we show that if a timelike or null geodesic hits the ring singularity 
 $(r=0, \vartheta=\frac {\pi}{2})$ then necessarily it is an
 equatorial geodesic.
 
 This important conclusion has been proven by Carter for the case of a Kerr (or Kerr-Newman) ring singularity
 and here we show that
 the inclusion of a positive $\Lambda$
 does not spoil this property.
 In order to prove this property, at first
 we construct the other two components
 of any causal geodesics
 relative to an arbitrary Carters block by making use of eqs. (\ref{Eq:GTM}, \ref{Eq:GPM}).\\
 
 If $r(\lambda), \vartheta(\lambda)$ 
 is any solutions of  $(\ref{Eq:GR},\ref{Eq:GTH})$, then by integrating 
(\ref{Eq:GTM}, \ref{Eq:GPM})
along  this solution, we obtain: 
\begin{equation}
 t(\lambda)-t(\lambda_{0})= \int_{\lambda_{0}}^{\lambda}\left[ \frac {I^{2}(r^{2}+a^2)}{\rho^{2}\Delta(r)}[(r^{2}+a^{2})E-al_{z}]+\frac {I^{2}a}{\rho^{2}{\hat \Delta}(\vartheta)}[l_{z}-aE sin^{2}\vartheta]\right] d\lambda'= \nonumber
\end{equation}
\begin{equation}
 =\pm \int_{r_{0}}^{r}\frac {I^{2}(r^{2}+a^2)}{\Delta(r)}[(r^{2}+a^{2})E-al_{z}]\frac {dr}{\sqrt{R(r)}}
\pm\int_{\vartheta_{0}}^{\vartheta}\frac {I^{2}a}{\hat {\Delta}(\vartheta)}[l_{z}-aE sin^{2}\vartheta]
\frac {d\vartheta} { \sqrt{\Theta(\vartheta)} }
\label{Eq:IGTM}
\end{equation}

\begin{equation}
\varphi(\lambda) -\varphi(\lambda_{0})=\int_{\lambda_{0}}^{\lambda}\left[ \frac {I^{2}a}{\rho^{2}\Delta(r)}[(r^{2}+a^{2})E-al_{z}] 
 - \frac {I^{2}a}{\rho^{2}{\hat \Delta}(\vartheta)}[E-\frac {l_{z}}{sin^{2}\vartheta}]\right] d\lambda'= \nonumber
\end{equation}
\begin{equation}
=\pm \int_{r_{0}}^{r}\frac {I^{2}a}{\Delta(r)}[(r^{2}+a^{2})E-al_{z}]\frac {dr}{\sqrt{R(r)}} 
\mp\int_{\vartheta_{0}}^{\vartheta}\frac {I^{2}a} {\hat {\Delta}(\vartheta)} [E-\frac {l_{z}}{sin^{2}\vartheta}]\frac {d\vartheta} { \sqrt{\Theta(\vartheta)} }.
 \label{Eq:IGPM}
\end{equation}
where we used
the property that the affine parameter $\lambda$
along $r(\lambda), \vartheta(\lambda)$ 
obeys:
 \begin{equation}
 \frac {d\lambda}{\rho^{2}}=\pm  \frac {dr}{\sqrt{R(r)}}=\pm  \frac {d\vartheta}{\sqrt{\Theta(\vartheta)}} 
  \label{Eq:APSA}
\end{equation}
Setting aside for the moment problems of convergence,
the 
right hand side of (\ref{Eq:IGTM},\ref{Eq:IGPM}),
provide a (formal)  solution of 
(\ref{Eq:GTM}, \ref{Eq:GPM})
once a solutions of  $(\ref{Eq:GR}-\ref{Eq:GTH})$
is a-priori known.\\

In order to show that only particular families
of equatorial geodesics can reach the
 ring singularity,
it is sufficient to consider geodesics on the  block $(T,g)$
that contains the ring singularity ie the block specified 
by the condition that $r\in (r_{1}, r_{2})$.
Let $(t(\lambda), r(\lambda),\vartheta(\lambda),\varphi(\lambda))$
represents an arbitrary (possibly a segment of a) timelike or a null geodesic within this 
 block, that hits the ring singularity.
 This means that there exist an $[\lambda_{0}, \lambda_{1})$
 so that $lim_{\lambda\to \lambda_{1}} r(\lambda)=0$ 
 and $lim_{\lambda\to \lambda_{1}} \vartheta(\lambda)=\frac {\pi}{2}$. 
 These condition imply  $R(r(\lambda))\geq 0$ and
 $\Theta(\vartheta(\lambda))\geq0$ 
 and continuity arguments require
 $R(0)\geq 0$ and $\Theta(\frac {\pi}{2})\geq0$.
 However a look at (\ref{Eq:THN}) and
  (\ref{Eq:MR}) shows that 
  $R(0)=-a^{2}Q$ and $\Theta(\frac {\pi}{2})=Q$
  and a  compatibility between these two conditions is obtained by the choice: $Q=0$. 
Since for $Q=0$, the condition $\Theta(\frac {\pi}{2})=0$  
 implies that  $\frac {d\vartheta}{d\lambda}=0$ at $\vartheta=\frac {\pi}{2} $,  
 then the totally geodesic property of the equatorial plane,
implies  that this geodesic is indeed an equatorial geodesic.
Thus only equatorial causal geodesics can reach the ring singularity
and in the next section we show
 that only finite amount of the affine parameter $\lambda$
 is required  by these geodesics to reach the ring singularity.\\
 
The capture of equatorial geodesics by the ring singularity raises the interesting issue
of characterizing  these geodesics.
All equatorial geodesics have $Q=0$, but it is not true that all geodesics with $Q=0$
are necessarily equatorial ones as can be easily seen from the structure of 
$(\ref{Eq:GR}-\ref{Eq:GTH})$. Moreover not all equatorial geodesics are captured by the 
singularity.
Does therefore,  exist a relation (or relations) upon  $(E,  l_{z}, m \geq 0)$ 
which guarantees capture of the geodesic by the ring singularity? 
Do the set of captured  geodesics constitute a set of ``measure zero` or they define an open set on the 
$(E,  l_{z}, m \geq 0)$ parameter space?
As  far as we are aware, these  problems are open even for the case of the Kerr singularity
and thus providing an answer would be a worthwhile project.\\

We finish this section by discussing particular 
families of causal geodesics 
that are natural consequences
of the special structure of 
 $(\ref{Eq:GR}-\ref{Eq:GTH})$. At first, one may 
 look for geodesics that satisfy $r(\lambda)=r_{0}$ for all $\lambda$, ie geodesics that are confined to
 move on an $r=r_{0}$ coordinate surface. Clearly, if they exist, they  satisfy:
 \begin{equation}
R(r_{0})=\frac {dR(r_{0})}{dr}=0,\quad\quad \Theta(\vartheta)\geq0,\quad \vartheta \in [a,b]\subset [0,\pi]
 \label{Eq.S1}
\end{equation}  
The condition $R(r_{0})=0$ combined with
$R(r)=I^{2}P^{2}(r)-{\Delta(r)}(m^{2}r^{2}+l^{2})$ imply that such geodesics (often referred as orbits) if they exist, 
they lie on blocks where
${\Delta(r)}>0$. Restricting $r$ in such domains, then the three conditions in (\ref{Eq.S1}) should determine $r_{0}$
and place restrictions  upon the constants $(l^{2}, E,  l_{z}, m \geq 0)$ for the ocurrene of such orbits.
Closely related to spherical orbits are polar orbits  or polar spherical orbits. These again are geodesics 
having the property that are crossing the north and south part of rotation axis 
at least once while  spherical polar orbits are
restricted to move at a fixed $r_{0}$ coordinate surface
and thus cross the rotation axis infinitely many times. Conditions for  the occurrence of spherical polar
orbits are described by
$(\ref{Eq.S1})$  except that for these orbits  $\Theta(\vartheta)$ is required to be positive on the entire interval $[0,\pi].$ 
Conditions for the occurrence of polar orbits are not so straightforward to state
and in the Appendix we discuss 
some of the problems associated with the existence of polar orbits.
It is worth pointing out
that for a Kerr background, 
 both of these type of orbits 
 exists
 and we expect  to 
occur also for the case of a Kerr-de Sitter, although we are not aware of any  result towards this direction
(some pertinent comments about spherical polar orbits have been made in ref \cite{Laz}).
In the last section of this paper, we prove the existence of causal geodesics that cross the rotation axis
but beyond this assertion existence of the geodesics mentioned above needs to proven.\\

Finally we discuss briefly causal geodesics 
characterized by $E=l_{z}=0$. From
$R(r)=I^{2}P^{2}(r)-{\Delta(r)}(m^{2}r^{2}+l^{2})$ it follows that such geodesics if exist,
are confined  on blocks where $\Delta(r)<0$. Returning to equations
(\ref{Eq:GTM}, \ref{Eq:GPM}) and setting $E=l_{z}=0$, it follows that these
geodesics are lying on the $(t=t_{0}, \varphi=\varphi_{0})$ two-surface referred as a polar plane
through  $(t_{0},\varphi_{0})$.
Since on any  block with $\Delta(r)<0$, the coordinate field $\frac {\partial}{\partial r}$
is timelike, these geodesics are necessary timelike or null.
They can be thought as the analogue of  the timelike geodesics sneaking from the white hole region 
into the black hole region
via the bifurcation two sphere associated with the intersection of the past and future event horizon of a Schwarzschild  black hole.
In the section $V$ we discuss 
further properties of Kerr-de Sitter spacetimes 
where the analogues to the Schwarzschild case would  become clearer.

\section{On the Completeness of Carter's-Blocks}
In this section,
 we study 
solutions of $(\ref{Eq:GR}-\ref{Eq:GTH})$ 
by first formulating  a suitable IVP.
For this, we begin with an arbitrary block $(T,g)$ and choose
a point $q=(t_{0}, \varphi_{0}, r_{0},\vartheta_{0})$ within
this block subject to the restrictions: $sin\vartheta_{0}\neq0$ and $\rho^{2}(r_{0},\vartheta_{0})>0$ ie
$q$ it is not part of the rotation axis\footnote{By the definition of a Carter's block,
note that always $\Delta(r_{0})\neq 0$.} neither
  lies on the ring singularity. Moreover we choose the constants\footnote{For the following analysis,
 we keep them fixed except
 whenever it is stated explicitly.} 
 $(l^{2},E,  l_{z}, m \geq 0)$ so that 
 $R(r_{0})>0$ and $\Theta(\vartheta_{0})>0$. 
With these choices, there exists an open, connected set $D$ in the $(r,\vartheta)$
plane containing $(r_{0},\vartheta_{0})$
and so that  on this $D$, $(R, \Theta)$ are bounded, and strictly positive.
Strict positivity of $(R, \Theta)$, implies that  
 $\sqrt {R}$ and $\sqrt{\Theta}$ 
 are continuously differentiable on $D$. The precise form of the domain $D$ 
 depends upon the nature of the block under consideration and upon the constants
  $(l^{2},E,  l_{z}, m \geq 0)$.
  The associated boundary
 $\partial D$ of $D$  is defined as the set of points\footnote{Notice that boundary points may include
 points $r_{i}$ where $\Delta(r_{i})=0$ but $R(r_{i})$ is non vanishing. The reason for this inclusion is explained in the text.} where $R(r)=0$
 or $\Theta(\vartheta)=0$.
 Both $D$ and $\partial D$
 play an important role in the  continuation of the solutions
of $(\ref{Eq:GR}-\ref{Eq:GTH})$
and for this reason 
we examine in details their properties.\\

The assumption that $q$ is off the rotation axis combined with
 $\Theta(\vartheta_{0})>0$,
imply  that at least when $l_{z}\neq0$, always  exist an interval $(\vartheta_{1},\vartheta_{2})\subset (0,\pi)$ so  that $\Theta(\vartheta_{1})=\Theta(\vartheta_{2})=0$  
 and $\Theta(\vartheta)>0$ in $(\vartheta_{1},\vartheta_{2})$.
Besides the interval $(\vartheta_{1},\vartheta_{2})$, we need to specify an open interval 
around $r_{0}$ so that $R(r)$ is positive. Once
 these intervals have been specified,
 their cartesian product\footnote{For this section, the plane $R^{2}$ is equipped
 with the product topology rather the more familiar metric topology arising from the Euclidean metric of the plane.
 Even though these two topologies are equivalent, the former is more suitable for establishing existence and uniqueness of IVP,
for details, see for instance ref. \cite{ODES}.}
defines the domain $D$. \\
 In this section, we take
$(\vartheta_{1},\vartheta_{2})\subset (0,\pi)$ either by assuming $l_{z}\neq0$ or if  $l_{z}=0$ by restricting the remaining 
$(l^{2},E, m \geq 0)$ in a manner
that 
 $(\vartheta_{1},\vartheta_{2})\subset (0,\pi)$. In the Appendix II, we deal with the special case
 where the constants 
 $(l^{2},E, m \geq 0)$ 
 have been chosen so that $\Theta(\vartheta)>0$ on the  entire interval $[0,\pi]$.\\

Under these conditions outlined above, we  first consider an interior block 
so that  $\Delta(r)<0$ in $(r_{i}, r_{i+1})$. 
If $P(r_{i})\neq0$
 and $P(r_{i+1})\neq0$
 then necessarily $R(r)>0$ on  $ [r_{i}, r_{i+1}]$
 and thus  for any $r_{0} \in (r_{i}, r_{i+1})$
and $\vartheta_{0}\in (\vartheta_{1},\vartheta_{2})$,
the domain $D$ and $\partial D$ have the form: 

\begin{equation}
D=\left\{ (r, \vartheta), r \in (r_{i}, r_{i+1}),
  \vartheta\in (\vartheta_{1},\vartheta_{2}) \right\},
 \label{Eq:D1}
\end{equation}  

 \begin{equation}
\partial D=\left\{ r=r_{i}, \vartheta_{1}\leq \vartheta\leq \vartheta_{2}\right\}\cup
\left\{\vartheta=\vartheta_{2}, r_{i}\leq r \leq r_{i+1}\right\}\cup\left\{r=r_{i+1}, \vartheta_{1}\leq \vartheta\leq \vartheta_{2}\right\}\cup\left\{
   \vartheta=\vartheta_{1}, r_{i}\leq r\leq r_{i+1}\right\}    
\label{Eq:BN}
\end{equation} 
Here,  even  though $R(r_{i})>0$ and $R(r_{i+1})>0$,
 we include $\left\{ r=r_{i}, \vartheta_{1}\leq \vartheta\leq \vartheta_{2}\right\}$
and $\left\{r=r_{i+1}, \vartheta_{1}\leq \vartheta\leq \vartheta_{2}\right\}$
as part of $\partial D$ since
equations $(\ref{Eq:GR}-\ref{Eq:GTH})$ 
have been derived
relative to a Boyer-Lindquist coordinates
where  $\Delta(r)$ is non vanishing\footnote{ In the present  work, we 
treat the system (\ref{Eq:GTM}-\ref{Eq:GTH})
as defined only relative to a specific block $(T,g)$
 and in particularly the ``radial`` like eq.
 (\ref{Eq:GR}) valid for $r\in (r_{i}, r_{i+1})$. Many of the geodesics
 reach the boundary of the block ie reach Killing horizons and 
 they must be continued through these horizons
 and one of the purposes of this work, is to discuss in details this continuation process.
 In the interesting approach of ref. \cite{Laz} the emphasis has been
  restricted to causal geodesics confined on a specific Carters block, 
 and the  analysis in \cite{Laz} contains many fine details of geodesic motion
 in both Kerr-de Sitter and Kerr-anti de Sitter.}.\\

For the case where $P(r_{i})=0$, $P(r_{i+1})\neq0$
then $R(r_{i})=0$ 
and $R(r_{i+1})\neq0$ and thus
 $D$ and $\partial D$ are identical to those in (\ref{Eq:D1}, \ref{Eq:BN}) except that
$R(r_{i})=0$ on $\left\{ r=r_{i}, \vartheta_{1}\leq \vartheta\leq \vartheta_{2}\right\}$.
The structure of $D$ and $\partial D$ for the cases
where
$P(r_{i})\neq0$ but $P(r_{i+1})=0$ 
 or  $P(r_{i})=P(r_{i+1})=0$
 which imply $R(r_{i})=R(r_{i+1})=0$, 
 can be easily inferred from 
 (\ref{Eq:D1}, \ref{Eq:BN}).\\\

It should be mentioned that for $\Lambda>0$ and for the case where 
 $\Delta (r)=0$ admits four distinct real roots
$r_{1}<0<r_{2}<r_{3}<r_{4}$,
then  $\Delta (r)<0$ for $r\in (r_{2},r_{3})$.
Moreover, from 
$R(r)=I^{2}P^{2}(r)-{\Delta(r)}(m^{2}r^{2}+l^{2})$, it  is not difficult to show that there exist parameters $(E,l_{z})$
where all four different boundaries $\partial D$ can be realized.
However, if the parameters $(M, a,\Lambda)$ are chosen so that
the roots 
$r_{2}$ and $r_{3}$ coalesce, then the region where $\Delta (r)<0$ disappears
and thus one left with the two asymptotic blocks where
 $\Delta (r)<0$.\\\
 
 We now consider a block  
where $\Delta(r)>0$ on  $(r_{i}, r_{i+1})$
and here the enumeration of  all $D$ and $\partial D$
 becomes a more tedious job since $R(r)=0$
 may admit roots in $(r_{i}, r_{i+1})$.
Bellow, we discuss
 only those $D$, $\partial D$ that lead to different continuation modes for
  the solutions of the IVP defined by (\ref{Eq:IVPRT1}).
 
If  the equation $R(r)=0$ admits 
 two  roots $\hat{r}_{1}< \hat {r}_{2}$ in 
 the interior of $(r_{i}, r_{i+1})$ so that $R(r)>0$ on  $(\hat r_{1}, \hat r_{2})$,
then for any $r_{0}\in (\hat{r}_{1}, \hat {r}_{2})$ and $ \vartheta\in (\vartheta_{1},\vartheta_{2})$
we take $D$ and its boundary in the form:
    \begin{equation}
D=\left\{ (r, \vartheta), r \in (\hat r_{1}, \hat r_{2}),
  \vartheta\in (\vartheta_{1},\vartheta_{2}) \right\},
   \label{Eq:D2}
\end{equation}  
 \begin{equation}
\partial D=\left\{ r=\hat r_{1}, \vartheta_{1}\leq \vartheta\leq \vartheta_{2}\right\}\cup
\left\{\vartheta=\vartheta_{2}, \hat r_{1}\leq r \leq \hat r_{2}\right\}\cup\left\{r=\hat r_{2}, \vartheta_{1}\leq \vartheta\leq \vartheta_{2}\right\}\cup\left\{
   \vartheta=\vartheta_{1}, \hat r_{1}\leq r\leq \hat r_{2}\right\}    
 \label{Eq:B2}
\end{equation}

If $R(r)=0$ admits 
another pair of roots
$\hat r_{3}, \hat r_{4}$ within $(r_{i}, r_{i+1})$
then 
 $D$ and $\partial D$
are as in (\ref{Eq:D2}, \ref{Eq:B2}) with the exception 
 that $\hat r_{1}, \hat r_{2}$ are replaced
by $\hat r_{3}, \hat r_{4}$.
For the case where $R(r)$ is positive  on $[r_{i}, r_{i+1}]$,
 then
$D$ and $\partial D$ are identical as those in (\ref{Eq:D1}, \ref{Eq:BN}). (In Figs. (1,2) graphs of $R(r)$ and  $\Delta(r) $
with the intervals of positivity and roots are shown.)\\

For completeness,
we briefly discuss the case where 
$P(r_{i})=0$, $P(r_{i+1})\neq0$
and thus $R(r_{i})=0$ 
and $R(r_{i+1})\neq0$.
Since 
$\Delta(r)>0$ in $(r_{i}, r_{i+1})$,
there is the possibility
that $R(r)=0$
admits a root $\hat{r}_{1}$  in the interior of  
$(r_{i}, r_{i+1})$ so that
$R(r)>0$ on $(r_{i}, \hat r_{1})$.
In this case  $D$ and $\partial D$
are as in  
 (\ref{Eq:D2}) and  (\ref{Eq:B2})
with the exception that  $\hat{r}_{1}$ 
is replaced by $r_{i}$.
Notice that here $r_{i}$ is necessary a double root
of $\Delta(r)=0$ while  $\hat r_{1}$
can be a single of a higher multiplicity root of $R(r)=0$.
A variation of this 
setting  corresponds to the case where 
 $\hat r_{1}$ is double root of $R(r)=0$
 so that  $R(r)$ is positive on $(r_{i}, \hat r_{1})$
 and $( \hat r_{1}, r_{i+1})$ and in this case 
  the associated $D$ and $\partial D$ are easily inferred.\\

Finally we discuss the case of the asymptotic block 
 $( r_{4},\infty)$.
 For this block (and for $\Lambda> 0$), it follows 
 from 
 $R(r)=I^{2}P^{2}(r)-{\Delta(r)}(m^{2}r^{2}+l^{2})$
 that
 $R(r)$ is  positive  on $( r_{4},\infty)$ and thus
 $D$ and $\partial D$ have the form:
 \begin{equation}
D=\left\{ (r, \vartheta), r \in (r_{4}, \infty),
  \theta\in (\vartheta_{1},\vartheta_{2}) \right\} 
  \label{Eq:D3}
\end{equation}  
 \begin{equation}
\partial D=\left\{ r=r_{4}, \vartheta_{1}\leq \vartheta\leq \vartheta_{2}\right\}\cup
\left\{\vartheta=\vartheta_{2}, r_{4}\leq r <\infty\right\}\cup\left\{
   \vartheta=\vartheta_{1}, r_{4}\leq r <\infty\right\}    
\label{Eq:AB}
\end{equation} 
For the case where $(T,g)$ is the other asymptotic block ie  $(-\infty, r_{1} )$, 
 then $\partial D$ is as above except that $r \to \infty$ is replaced by $r\to -\infty$ and some inequalities are reversed.\\
   
With the above preliminary work
on the possible domains $D$, 
we now consider the (IVP):
 \begin{equation}
\rho^{2} \frac {dr}{d\lambda}= {R(r)}^{\frac{1}{2}},\quad \rho^{2} \frac {d\vartheta}{d\lambda}= {\Theta(\vartheta)}^{\frac{1}{2}},\quad r(\lambda_{0})=r_{0},\quad \vartheta(\lambda_{0})=\vartheta_{0},\quad (r_{0},\vartheta_{0}) \in D
  \label{Eq:IVPRT1}
\end{equation}
where $D$ is one of the domains defined above
and
at first we restrict attention to the positive sign\footnote{Had we have chosen 
a different combination of signs in (\ref{Eq:GR}-\ref{Eq:GTH})
the following analysis remains intact for these choices as well.}
in (\ref{Eq:GR}-\ref{Eq:GTH})
(however, as we shall see shortly
both signs  in (\ref{Eq:GR}-\ref{Eq:GTH})
play an important role in the continuation of maximal solutions of this IVP).\\

Since the functions $R$ and $\Theta$ are
non vanishing, bounded and continuously differentiable on any $D$ 
in (\ref{Eq:IVPRT1}), theorems on the existence, uniqueness and continuation of solutions 
of IVP's, affirm the existence of a unique  solution 
$r(\lambda), \vartheta(\lambda), \lambda \in [\lambda_{0}, \lambda_{1})$ with $ [\lambda_{0}, \lambda_{1})$
a maximal interval so that 
the $(lim_{\lambda\to\lambda_{1}}r(\lambda), lim_{\lambda\to\lambda_{1}}\vartheta(\lambda))$ 
exist and belong to $\partial D$
(see for instance \cite{ODES}, section $3$ and in particular Theorem $3.6$).
Of a crucial importance for the problem of geodesic completeness
is the nature of the interval
$ [\lambda_{0}, \lambda_{1})$.
If $\lambda_{1}\to \infty$, then  $(r(\lambda), \vartheta(\lambda))$
is an inextendible to the right  solution of (\ref{Eq:IVPRT1})
and for this solution, the system (\ref{Eq:IGTM},  \ref{Eq:IGPM}) can be integrated 
yielding an inextendible to the right  geodesic $(t(\lambda),\varphi(\lambda),r(\lambda),
\vartheta(\lambda))$  ie a geodesic defined for all $\lambda \in  [\lambda_{0}, \infty).$ 
By construction this part of the geodesic remains within the block\footnote{
Similar analysis holds for the solutions
of $(\ref{Eq:IVPRT1})$ defined on a maximal domain of the form: $(\hat \lambda_{1}, 
\lambda_{0}]$. For the case where $\hat \lambda_{1}\to -\infty$, the solution is left inextendible
and an integration of (\ref{Eq:IGTM},  \ref{Eq:IGPM}) along this solution
yields a left inextendible geodesic defined in $(T,g)$.
A geodesic is complete if it is simultaneously left and right inextendible. 
For simplicity, in the main text, we restrict our attention
to the maximal interval of the form: $[\lambda, \lambda_{1})$.} 
 $(T,g)$. If on the other hand 
 $\lambda_{1}< \infty$, then the solution
 $r(\lambda), \vartheta(\lambda), [\lambda_{0}, \lambda_{1}) $
can be continued into larger $\lambda$ intervals and the mode of this continuation will be discussed  
further ahead.\\

Whether $\lambda_{1}$
is bounded or unbounded
depends upon the multiplicity of the zeros of the functions $R(r)$ and $\Theta(\vartheta)$.
The role of the multiplicity 
of the zeros of $R(r)$ and $\Theta(\vartheta)$,
it can be seen 
by noting that along any solution  $r(\lambda), \vartheta(\lambda)$ of (\ref{Eq:IVPRT1}),
we have:
\begin{equation}
\lambda-\lambda_{0}= \int\limits_{r_{0}}^{r(\lambda)}\frac {r^{2}dr}{ \sqrt{{R(r)}}} +
 \int\limits_{\vartheta_{0}}^{\vartheta(\lambda)}\frac {a^{2}cos^{2}{\vartheta}d\vartheta}{ \sqrt {{\Theta(\vartheta)}}},\quad \lambda \in [\lambda_{0},  \lambda)
\label{Eq:APP}
\end{equation}
Therefore, if for  a particular solution
 $(r(\lambda), \vartheta(\lambda)), \lambda \in [\lambda_{0}, \lambda_{1})$ 
 it turns out that $\lim_{\lambda \to \lambda_{1}}r(\lambda):=\hat r_{1}<\infty$ or (-and) $\lim_{\lambda \to \lambda_{1}}\vartheta (\lambda):=\hat \vartheta_{1}$ 
 are simple root of $R(r)$, respectively $\Theta(\vartheta)$,
then (\ref{Eq:APP}) converges. If however either $\hat r_{1}$ or $\hat \vartheta_{1}$ are
 roots of higher multiplicity
 then $\lambda_{1} \to \infty$
 and in that event the solution approaches 
 the boundary points $\hat r_{1}$ or $\hat \vartheta_{1}$ 
  only asymptotically ie as $\lambda\to \infty$
 and clearly  no continuation of the solution is required.\\
 
 It is worth mentioning
 that for the case where $(T,g)$
 is one of the asymptotic blocks,
 say the block specified  by: $r>r_{4}$ and the solution
  $(r(\lambda), \vartheta(\lambda))$ 
  runs into the asymptotic region, ie  
 $\lim_{\lambda \to \lambda_{1}}r(\lambda) \to +\infty$, then (\ref{Eq:APP}) 
 combined with the asymptotic form of $R(r)$
shows that for timelike geodesics
necessary   $\lambda_{1} \to \infty$ and the divergence is logarithmic,
  while for the case of null geodesics $\lambda_{1} \to \infty$   
  and the integral in (\ref{Eq:APP}) diverges linearly.\\\

We now discuss in details
the continuation of maximal solutions
$r(\lambda), \vartheta(\lambda), \lambda \in [\lambda_{0}, \lambda_{1})$ 
of (\ref{Eq:IVPRT1}) for the case where the domain $D$ is described by (\ref{Eq:D1}).
Since for this $D$, 
$R(r)>0$ on $[r_{i},r_{i+1}]$,
any maximal solution 
either reaches a point on $\left\{r=r_{i+1}, \vartheta_{1}\leq \vartheta\leq \vartheta_{2}\right\}$
 or reaches a point on 
$\left\{\vartheta=\vartheta_{2}, r_{i}\leq r \leq r_{i+1}\right\}$
as  $\lambda \to \lambda_{1}$.
If the first possibility occurs, then $\lambda_{1}$ is necessary
 finite, and the solution 
 reaches the boundary of the  block 
  in the sense  
   $\lim_{\lambda \to \lambda_{1}}\Delta (r(\lambda))=\Delta(r_{i+1})=0$.
 However this situation does not generates geodesic incompletness
 since
 the solution can be continued as   
 a solution of (\ref{Eq:GR}-\ref{Eq:GTH})
 into an 
 adjacent block.
in order to pursue this continuation,  we need to embed the Carters block
into a larger spacetime domains
and  technicalities of this embedding will be discussed in the next section.\\
 
 If the second possibility occurs, ie the solution reaches
 a point 
on  $\left\{\vartheta=\vartheta_{2}, r_{i}\leq r \leq r_{i+1}\right\}$ 
 for $\lambda_{1}< \infty$,
 then since $\frac {d\vartheta}{d\lambda}=0$ at this part of the boundary,
the solution
can be continued in a smooth manner
through
the point $(\hat r_{0}, \vartheta_{2})\in \partial D$
as a solution of the IVP: 
\begin{equation}
\rho^{2} \frac {dr}{d\lambda}= {R(r)}^{\frac{1}{2}},\quad \rho^{2} \frac {d\vartheta}{d\lambda}= -{\Theta(\vartheta)}^{\frac{1}{2}},\quad r( \lambda_{1})=\hat r_{0},\quad \vartheta(\lambda_{1})=\vartheta_{2}
 \label{Eq:NIVP}
\end{equation}
Since $\Theta(\vartheta_{2})=0$, it is not any longer true that
both $ {R(r)}^{\frac{1}{2}}, {\Theta(\vartheta)}^{\frac{1}{2}}$ 
are continuously differentiable  
in an open vicinity of the initial point:
$(r( \lambda_{1})=\hat r_{0}, \vartheta(\lambda_{1})=\vartheta_{2})$
and thus the existence and uniqueness of a solution 
of (\ref{Eq:NIVP}) needs to be looked upon carefully.
As it turns out, the fact that 
${\Theta(\vartheta)}$ has a simple zero at $\vartheta_{2}$
permits us to prove an existence and uniqueness 
property of the 
 IVP in (\ref{Eq:NIVP})
 and all details calculations leading to this conclusion are 
 discussed in the Appendix $I$.\\

Based on the results of this Appendix, 
there exist a maximal interval of existence $[\lambda_{1},  \lambda_{2})$,
where the solution of
(\ref{Eq:NIVP}) is defined in $D$ and it is monotonically decreasing.
Here again
 $\lim_{\lambda \to \lambda_{2}}(r(\lambda),\vartheta(\lambda))$,  
 is a boundary point
and depending upon the nature of this new boundary point, the solution either can be continued in an adjacent block or
will be reflected once a point
on $\left\{\vartheta=\vartheta_{1}, r_{i}\leq r \leq r_{i+1}\right\}$ is reached.\\

Notice that in this continuation process, we assumed
that $\vartheta_{2}$ and $\vartheta_{1}$ are   simple zeros of $\Theta(\vartheta)$. If
this assumption fails ie $\Theta(\vartheta)$ has double or higher order zero at 
$\vartheta_{2}$ (or and at $\vartheta_{1}$), then the boundary point
$\vartheta_{2}$  is approached only asymptotically ie as $\lambda \to \infty$. Since however 
only a finite amount of affine parameter 
is required for the solution to reach a point on 
$\left\{r=r_{i+1}, \vartheta_{1}\leq \vartheta\leq \vartheta_{2}\right\}$,
 therefore the solution  leaves the block under consideration
and thus a point on  $\left\{\vartheta=\vartheta_{2}, r_{i}\leq r \leq r_{i+1}\right\}$ 
cannot be reached by a solution of 
(\ref{Eq:IVPRT1}) with an initial point $(r_{0},\vartheta_{0})$. In Fig.3. we sketch the behavior of solutions (\ref{Eq:GR}-\ref{Eq:GTH}) 
for two cases of initial conditions.\\

We now discuss the continuation of solutions 
of the IVP (\ref{Eq:IVPRT1}) for the case where $D$ and $\partial D$ 
are still described by (\ref{Eq:D1}) resp.
(\ref{Eq:BN}) but now 
$P(r)$ and $\Delta(r)$ share a common root.
If we assume that this root is $r_{i+1}$
and moreover  it is a 
simple root of $\Delta(r)=0$ and thus a 
simple root
of $R(r)=0$, then again the solution 
$r(\lambda), \vartheta(\lambda), \lambda \in [\lambda_{0}, \lambda_{1})$ 
 either reaches a point on  $\left\{\vartheta=\vartheta_{2}, r_{i}\leq r \leq  r_{i+1}\right\} $
or reaches a point on $ \left\{ r= r_{i+1}, \vartheta_{1}\leq \vartheta\leq \vartheta_{2}\right\}$.
 For the case where
 $\vartheta_{2}$ is a simple zero of $\Theta(\vartheta)$,
 the solution
can be continued by employing the same IVP as in (\ref{Eq:NIVP})
coupled with the analysis of the Appendix $I$, while for  the case where
the solution reaches a point on $ \left\{ r=r_{i+1}, \vartheta_{1}\leq \vartheta\leq \vartheta_{2}\right\}$
and  even though $R(r_{i+1})=0$,
 the solution does not exhibit a turning point
 at $r_{i+1}$.
 This failure,
 is due to the fact 
$\Delta(r_{i+1})=0$
and clearly $r_{i+1}$
does not belong to the block  under consideration.
However, the solution 
 can be continued into adjacent block
and details of this continuation will be discussed in the next section.
\\
For the case where $r_{i+1}$ is 
a double or higher order root of $\Delta(r)=0$ 
then 
$r(\lambda), \vartheta(\lambda), \lambda \in [\lambda_{0}, \lambda_{1})$ 
 requires an infinite amount
 of affine parameter to 
reach a point on $ \left\{ r=r_{i+1}, \vartheta_{1}\leq \vartheta\leq \vartheta_{2}\right\}$
and thus
no further continuation is required\footnote{
Ought to be mentioned, 
if the solution reaches
 a point on $ \left\{ r= r_{i}, \vartheta_{1}\leq \vartheta\leq \vartheta_{2}\right\}$
since $R(r_{i})>0$, 
then a continuation into an adjacent block  is required.}.
  Although here we discussed the case where
 $P(r_{i+1})=0$ and $P(r_{i})\neq 0$
 similar  continuations modes for the maximal solutions
 of
 (\ref{Eq:IVPRT1})
 hold  for the case where $D$ and $\partial D$  
 correspond to the 
 case 
 $P(r_{i})=0$ and $P(r_{i+1})\neq0$ 
 or  $P(r_{i+1})=P(r_{i})=0$
 and we shall not discuss them any further.\\\

We now consider the  continuations of maximal solutions 
of (\ref{Eq:IVPRT1}) for the case
where $D$ and $\partial D$ are described by (\ref{Eq:D2}) resp.  (\ref{Eq:B2}). 
 For this case, any maximal solution  $(r(\lambda), \vartheta(\lambda)),  [\lambda_{0}, \lambda_{1})$, 
 either  reaches a point on  $\left\{\vartheta=\vartheta_{2}, \hat r_{1}\leq r \leq \hat r_{2}\right\} $
 or reaches a point on $ \left\{ r=\hat r_{2}, \vartheta_{1}\leq \vartheta\leq \vartheta_{2}\right\}$.
 If the first possibility occurs and  $\lambda_{1}< \infty$, 
 the solution can be continued as a solution of the IVP defined in (\ref{Eq:NIVP})
coupled with the results of Appendix $I$, 
while for the second possibility,
and as long as 
$\lambda_{1}< \infty$, the solution
 can be continued as a $C^{1}$ solution of the  IVP:
 \begin{equation}
\rho^{2} \frac {dr}{d\lambda}= -{R(r)}^{\frac{1}{2}},\quad \rho^{2} \frac {d\vartheta}{d\lambda}= {\Theta(\vartheta)}^{\frac{1}{2}},\quad r(\lambda_{1})=\hat r_{2},\quad \vartheta(\lambda_{1})=\vartheta
 \label{Eq:NNIVP}
\end{equation}
Structurally this IVP exhibits the same deficiencies as 
the IVP in  (\ref{Eq:NIVP})
with the function
$R(r)$ playing the same role as 
 $\Theta(\vartheta)$ played 
 in (\ref{Eq:NIVP}).
 The results of the Appendix $I$ are also applicable for this IVP.
 Accordingly, there
exist a unique solution defined on a maximal interval
$[\lambda_{1}, \lambda_{2})$.
As long as $\hat r_{1}, \hat r_{2}$ and $\vartheta_{1}, \vartheta_{2}$
are single roots of $R(r)=0$ resp. $\Theta(\vartheta)=0$,
then
$(r(\lambda), \vartheta(\lambda)),  [\lambda_{0}, \lambda_{1})$
 (and of  
 all solutions
(\ref{Eq:IVPRT1}))
get trapped within $D$ via multiple reflection upon $\partial  D$.
In Fig.4 a sketch of  solutions curves
and their continuations through the boundary are indicated.\\

In the event that some of the roots 
$\hat r_{1}, \hat r_{2}$ or $\vartheta_{1}, \vartheta_{2}$ are double or of higher 
multiplicity, still
the solutions is trapped. If for instance $\hat r_{1}, \hat r_{2}$
are simple zeros and 
$\vartheta_{2}$ is a double zero, then the solution
after multipole reflections at $\hat r_{1}$ and $\hat r_{2}$
reaches asymptotically ie as $\lambda\to \infty$ 
approaches a
point $\vartheta_{2}$ on the boundary
$\left\{\vartheta=\vartheta_{2}, \hat r_{1}\leq r \leq \hat r_{2}\right\}.$
A similar scenario holds if for instance 
$\hat r_{2}$ is a double zero and 
 $\vartheta_{1}, \vartheta_{2}$ are simple zeros.\\

We do not discuss the continuation processes for
  the case where 
 $\Delta(r)$   and $P(r)$ share a common root
 since the continuation of solutions  are of the form already discussed earlier on,
but rather we concentrate on  the continuation of solutions 
of  (\ref{Eq:IVPRT1})
for the case where $D$ 
and $\partial D$ are described
 by (\ref{Eq:D3})  resp. by (\ref{Eq:AB}) 
 ie are defined in the asymptotic block
 $(T,g)$ 
 specified by $r\in [r_{4}, \infty)$. Since $R(r)$ is positive definite on $D$
 any maximal solution
 $(r(\lambda), \vartheta(\lambda)),  [\lambda_{0}, \lambda_{1})$    
 that reaches the asymptotic region $r \to \infty$ 
 it does  so asymptotically ie 
  $\lambda_{1}\to \infty$.
Moreover solution of  (\ref{Eq:IVPRT1}) 
can either  reach  the asymptotic region  after
reflections 
 upon the boundary $ \left\{
   \vartheta=\vartheta_{1}, r_{4}\leq r <\infty\right\}$
 or upon 
$\left\{\vartheta=\vartheta_{2}, r_{4}\leq r <\infty\right\}$
or return to the point on the $r=r_{4}$ boundary. In Fig. 5, a solutions curves
defined on the asymptotic block specified by: $r\in [r_{4}, \infty)$,are plotted. \\   
 
 \begin{figure}
	\includegraphics[width=0.5\textwidth]{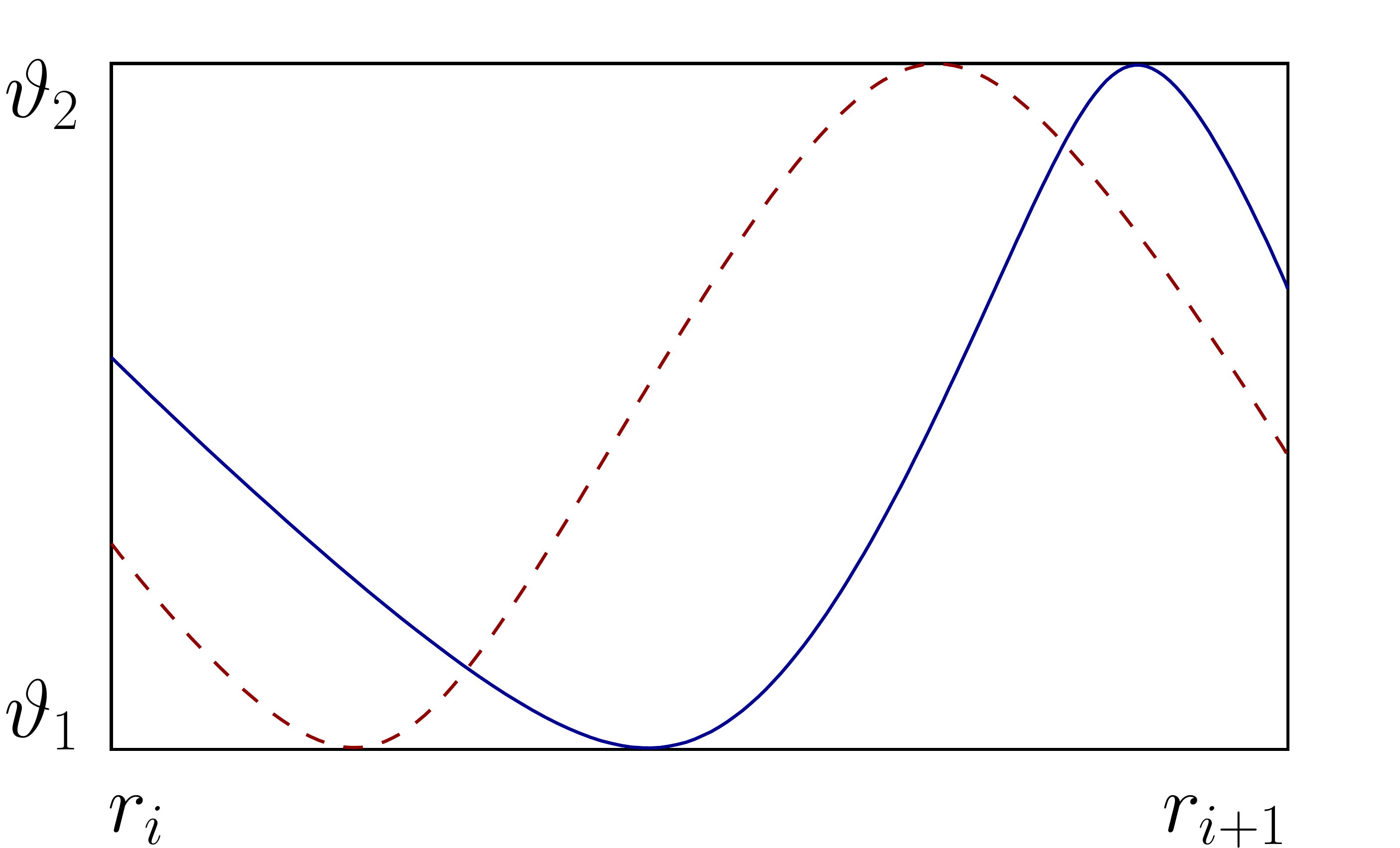}
	\caption{In this figure, two solution curves
	of (\ref{Eq:IVPRT1}) 	are shown and both solutions
	have been extended to the left as well.
	The solutions have a turning point
	at $\vartheta_{1}$ respectively $\vartheta_{2}$ part of the $\partial D$
	are continued until they  reach the $r_{i}$, $r_{i+1}$
	part of the $\partial D$.}
\end{figure}

\begin{figure}
	\includegraphics[width=0.5\textwidth]{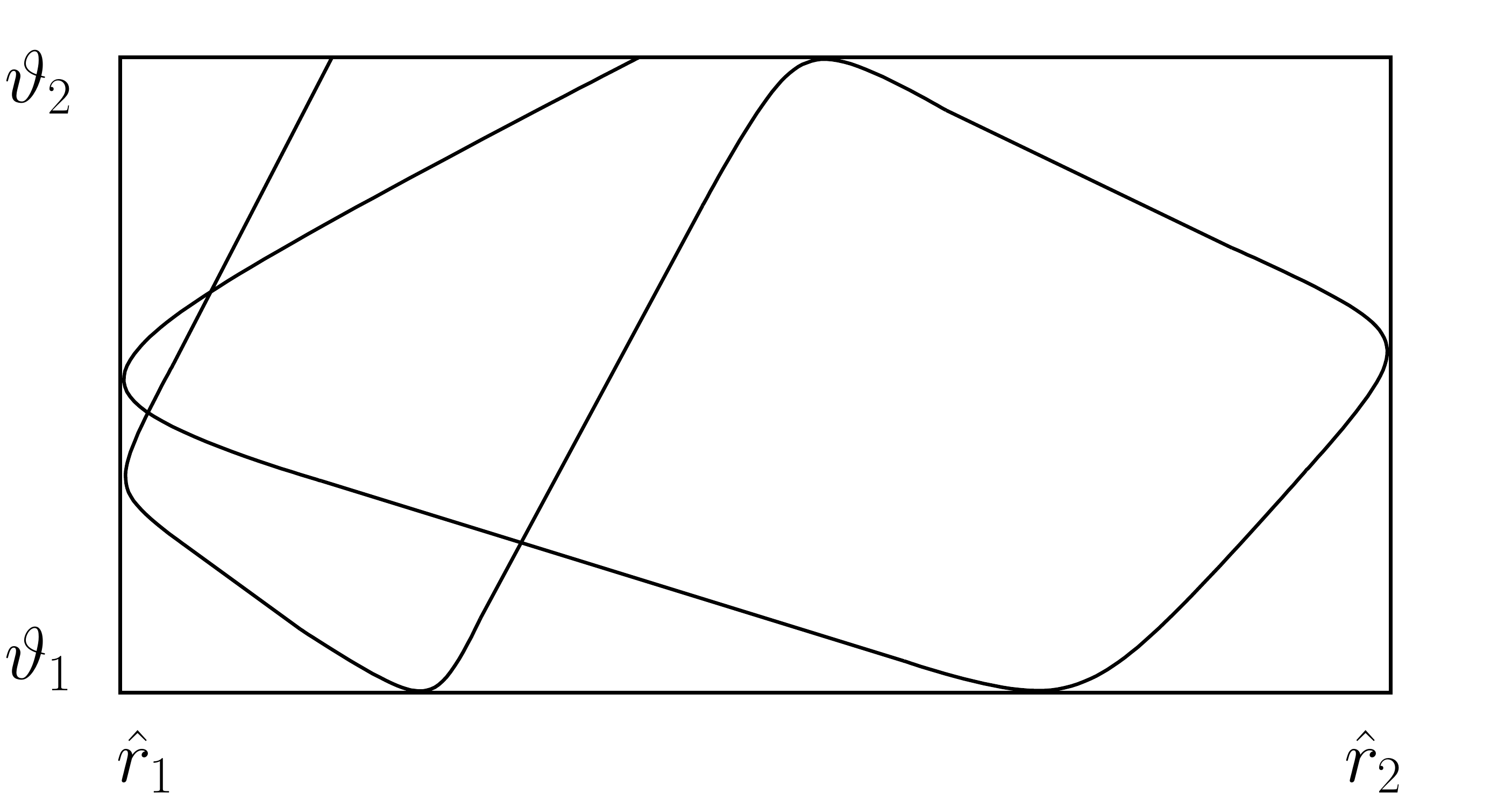}
	\caption{In this figure, a trapped solution curve
	is indicated. There exist turning points on each component
	of $\partial D$. Even it appears that the solution terminate at the $\vartheta_{2}$
	part of the boundary actually they have turning points and their motion continues for ever
	via reflections on the boundaries.}
\end{figure}

\begin{figure}
	\includegraphics[width=0.6\textwidth]{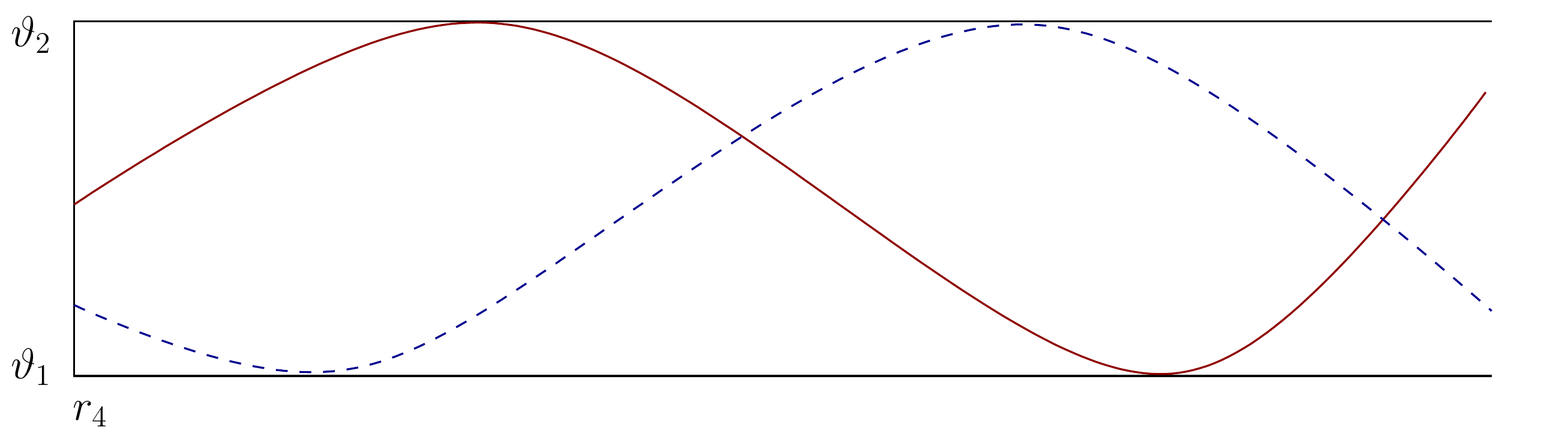}
	\caption{In this figure, two solution curves in $r \geq r_{4}$ block is
	indicated. There exist two turning points for each solution on each component
	of $\partial D$ and the solution continue in an interior adjacent block.}
	
\end{figure}

Finally  we consider the 
 block $(T, g)$ that harbor the ring singularity $(r=0, \vartheta=\frac { \pi}{2})$.
 As we have already seen,
 only equatorial causal geodesics reach
 points on the ring singularity. Here we show that such geodesics
 reach the ring singularity within a finite amount of the affine parameter.
 For that, let
 ($r(\lambda), \vartheta(\lambda)=\frac {\pi}{2}), \lambda \in [\lambda_{0}, \hat \lambda_{1})$
a solution of (\ref{Eq:IVPRT1})
having the property 
$\lim_{\lambda \to \lambda_{1}}\rho^{2}(r(\lambda), \vartheta(\lambda))\to 0$
i.e. the solution  hits the ring singularity. 
For this solution (\ref{Eq:APP})
implies that  only a finite amount of $\lambda$ is required for the solution
to teach the ring singularity. Accordingly
 timelike or null geodesics that hit the ring singularity
 are necessary incomplete.\\\

The so far analysis shows 
that  for the various domains $D$ introduced earlier on,
the solutions of the IVP (\ref{Eq:IVPRT1})
either remain trapped within the block $(T,g)$
for all $\lambda \in (-\infty, \infty)$
or they reach within  finite amount
of affine parameter the boundaries of the block, defined by $r=r_{i}$,
(or-and $r=r_{i+1}$).
In the first case, by integrating 
(\ref{Eq:GTM}, \ref{Eq:GPM}) along
these solutions we concluded that
there exist complete causal geodesics
that remain within a Carters block
while for the second possibility 
 these solutions are required to be continued
 across Killing horizons and this issue will be discussed in the next sections.
For the case where the block harbors the ring singularity,
there exist solutions
of (\ref{Eq:APP})
that reach the ring singularity in a finite amount of affine parameter
and these geodesics are incomplete.
In the next section we show that geodesics that run into the boundaries
of a Carters block can be continued  further
so they become complete, modulo of course those that run into
the ring singularity.\\

\section{Null geodesics on Kerr de Sitter }
In this section, first we show that any Carter«s block can be ««embedded««
into a larger Kerr-de Sitter region so that the boundaries 
$r=r_{i}$  (or-and $r=r_{i+1}$)
 become Killing horizons.
 This can be done by introducing 
 ingoing (respectively outgoing) 
 Finkelstein coordinates and 
 these coordinates, like for the case of ordinary Kerr, are based on the  principal null geodesics
congruences admitted by a Kerr-de Sitter metric.\\ 

We recall
that for an arbitrary block $(T,g)$, the equations describing 
null geodesics are 
obtained from (\ref{Eq:GT}-\ref{Eq:GTH}) by setting $m=0$.
For the study of these geodesics,
we again employ the  
parameter $Q$ and introduce new parameters 
$(\eta, \xi)$ defined via:
\begin{equation}
\frac {l^{2}}{I^{2}E^{2}}=\frac {Q}{I^{2}E^{2}}+(\frac {l_{z}}{E}-a)^{2}=\eta+(\xi-a)^{2},\quad \eta=\frac {Q}{I^{2}E^{2}},\quad \xi=\frac {l_{z}}{E}
\label{Eq:DQ}
\end{equation}
so that for $m=0$ the functions $R(r)$ and $\Theta(\vartheta)$ 
take the  form:
\begin{equation}
R(r)=I^{2}E^{2} \left[(r^{2}+a^{2}-a\xi)^{2} -{\Delta(r)}(\eta+(\xi-a)^{2}\right]
\label{Eq:DRN}
\end{equation}
\begin{equation}
\Theta(\vartheta)=I^{2}E^{2} \left[{\hat {\Delta}(\vartheta)}(\eta+(\xi-a)^{2})-(\xi cosc\vartheta-asin\vartheta)^{2}\right],\quad  cosc\vartheta=\frac {1}{sin\vartheta}.
\label{Eq:DTN}
\end{equation}
We now choose the constants $(\eta, \xi)$ so that 
\begin{equation}
\eta+(\xi-a)^{2}=0,\quad\xi cosc\vartheta=asin\vartheta.
\label{Eq:PND}
\end{equation}
where the condition $\xi cosc\vartheta=asin\vartheta$ fixes the angle (or angles) $\vartheta$ 
so that $\Theta(\vartheta)=0$.
With  these choices, $R(r)=I^{2}E^{2}\rho^{4}$, and thus (\ref{Eq:GT}-\ref{Eq:GTH}) reduce to
\begin{equation}
 \frac {dt}{d\lambda}=\frac {I^{2}E(r^{2}+a^{2})}{{\Delta(r)}},\quad \frac {d\varphi}{d\lambda}=\frac {I^{2}Ea}{{\Delta(r)}},\quad      
 \frac {dr}{d\lambda}=\pm \arrowvert IE \arrowvert,\quad  \frac {d\vartheta}{d\lambda}=0
\label{Eq:RTPND}
\end{equation}
Using the freedom in the choice of $\lambda$, we absorb the factor $IE$ into the 
new affine parameter (still denote by $\lambda$) and choose
the negative sign in the ``radial  equation``. We thus lead into a simple set of 
equations:
\begin{equation}
 \frac {dt}{d\lambda}=\frac {I(r^{2}+a^{2})}{{\Delta(r)}},\quad \frac {d\varphi}{d\lambda}=\frac {Ia}{{\Delta(r)}},\quad      
 \frac {dr}{d\lambda}=-1\quad  \frac {d\vartheta}{d\lambda}=0
\label{Eq:RTPNDD}
\end{equation}
which can be integrated explicitly.
The solutions define a preferred family of ''ingoing'' null geodesics
referred as the the principal (ingoing)  family. 
 Using these geodesic congruence  we define ingoing  Finkelstein coordinates
 $(v, \overleftarrow{\varphi},r,\vartheta)$  via
 \begin{equation}
dv=dt +\frac {I(r^{2}+a^{2})}{{\Delta(r)}}dr,\quad d\overleftarrow{\varphi}=d\varphi+\frac {Ia}{{\Delta(r)}}dr
\label{Eq:INC}
\end{equation}
ie  $(v, \overleftarrow{\varphi})$ remain constant along members of the principal (ingoing)  family.
Transforming the Kerr-de Sitter metric  $g$ in (\ref{Eq:g}) to
these new coordinates  $(v, \overleftarrow{\varphi}, r,\vartheta)$, we get
\begin{equation}
g=-\frac {\Delta(r)-a^{2} {\hat {\Delta}({\vartheta})sin^{2}\vartheta}}{I^{2}\rho^{2}}dv^{2}+\frac {2}{I}dvdr-2 \frac {a}{I}sin^{2}\vartheta
d\overleftarrow{\varphi}dr-2\frac {asin^{2}\vartheta[(r^{2}+a^{2}){\hat \Delta({\vartheta})-{\Delta(r)}}]}
{I^{2}\rho^{2}}dvd\overleftarrow{\varphi}+\nonumber
\end{equation}
\begin{equation}
+\frac {{\rho^{2}}} {{\hat {\Delta}({\vartheta})}}d\vartheta^{2}+\frac {\hat {\Delta}({\vartheta})(r^{2}+a^{2})^{2}-\Delta(r)a^{2}sin^{2}\vartheta}{I^{2}\rho^{2}}
sin^{2}\vartheta d\overleftarrow{\varphi}^{2}.
\label{Eq:gINC}
\end{equation}
where now $g$ is regular  
across the zeros  of $\Delta(r)=0$. 
We let the coordinates $(v,r)$ to run over the entire real line
and define the extended Kerr-de Sitter metric
to be  described  
by (\ref{Eq:gINC}).
The resulting spacetime,
has  an irremovable ringlike curvature singularity at $\rho^{2}=r^{2}+a^{2}cos^{2}{\vartheta}=0$
and a removable coordinate singularity along the ««rotation axis««. Different Carters blocks can be embedded
into spacetime region covered by this ingoing coordinate and the process is similar for the case of Kerr
(for a thorough discussion regarding these embeddings consult: \cite{Neil}).
\\

In these ingoing  $(v, \overleftarrow{\varphi}, r, \vartheta)$ coordinates, the  Killing fields take the form $\xi_{t}=\frac {\partial}{\partial v},\xi_{\varphi}=\frac {\partial}{\partial \overleftarrow{\varphi}}$ 
and a computation based on $(\ref{Eq:INC},\ref{Eq:gINC})$, shows that 
\begin{equation}
g(\xi_{t},\xi_{t})=-\frac {\Delta(r)-a^{2} {\hat {\Delta}(\vartheta)sin^{2}\vartheta}}{I^{2}\rho^{2}},\quad
g^{2}(\xi_{t},\xi_{\varphi})-g(\xi_{t},\xi_{t})g(\xi_{\varphi},\xi_{\varphi})=\frac {\sin^{2}{\vartheta}}{I^{4}}\Delta (r){\hat \Delta} (\vartheta)\label{Eq:KI}
\end{equation}
and thus for $\Lambda>0$, it follows that $\xi_{t}$ is spacelike  as $r\to \pm \infty$. Since $\xi_{t}$ can be  rescaled, we remove this freedom demanding that 
in the limit of $t\to \infty$, the field $\xi_{t}$ reduces to the standard form a de Sitter like form.
As for the  case of the Kerr metric, an analysis of the invariant $g(\xi_{t},\xi_{t})=0$ shows  the existence of
non trivial ergospheres 
and 
we expect the existence of Killing horizons at the ''interior'' of these ergospheres.
In order to locate these Killing horizons, we consider 
the combination $\xi=\xi_{t}+\Omega(r,\vartheta)\xi_{\varphi}$ with $\Omega(r,\vartheta)$
a smooth function and require  $\xi$ to be timelike and future pointing.
These condition demand
\begin{equation}
\Omega_{-}(r,\vartheta)< \Omega (r,\vartheta)<\Omega_{+}(r,\vartheta),\quad \Omega_{\pm}(r,\vartheta)=
\frac {-g(\xi_{t},\xi_{\varphi})\pm [g^{2}(\xi_{t},\xi_{\varphi})-g(\xi_{t},\xi_{t})g(\xi_{\varphi},\xi_{\varphi})]^{\frac {1}{2}}}{g(\xi_{\varphi},\xi_{\varphi})}
\label{Eq:TIM}
\end{equation}
and thus causal future directed timelike fields of the form $\xi=\xi_{t}+\Omega(r,\vartheta)\xi_{\varphi}$
do not exist in the regions where $\Delta (r)<0$. 

We now show that the set of points: $(v, \overleftarrow{\varphi}, r=r_{i}, \vartheta)$ 
where  $r=r_{i}$ stand for the real roots of
 $\Delta (r_{i})=0$, are null hypersurfaces.  For this we note
 that the normal vector $N$ of any $r=const$
 hypersurface, has the form:
 
\begin{equation}
N={\hat g}^{\mu\nu}\delta_{\nu}^{r}\frac {\partial}{\partial x^{\mu}}={\hat g}^{\mu r}\frac {\partial}{\partial x^{\mu}},\quad
x^{\mu}=(v, \overleftarrow{\varphi}, r, \vartheta)
\label{Eq:NKH}
\end{equation}
where ${\hat g}^{\mu\nu}$  stand for the contravariant components of $g$ 
relative to  $(v, \overleftarrow{\varphi}, r=r_{i},\vartheta)$ coordinates. 
Noting that
 \begin{equation}
g(N,N)={\hat g}^{rr}=g^{rr}=\frac {\Delta(r)}{\rho^{2}}
\label{Eq:MNKH}
\end{equation}
the claim that any $r=r_{i}$ hypersurface is a null hypersurface follows\footnote{
The term $\frac {\Delta(r)}{\rho^{2}}$ is well defined over the entire 
domain of validity of the ingoing chart
and this coupled with the fact left hand side of 
(\ref{Eq:MNKH}) is an analytic function relative to ingoing 
coordinates shows 
that the claim does not based on Boyer-Lindquist coordinates.
The latter have been used only as an intermediate step.}.

For each real root 
$r_{i}$ of $\Delta(r)=0$, let the constants
\begin{equation}
\Omega_{i}=-\frac {g(\xi_{t},\xi_{\varphi})}{g(\xi_{\varphi},\xi_{\varphi})}=\frac {a}{r_{i}^{2}+a^{2}}.
\label{Eq:ROT}
\end{equation}
where in above evaluation at $r_{i}$ is understood.
Using these 
$\Omega_{i}$, let the Killing field
\begin{equation}
\hat \xi_{i}=\xi_{t}+\Omega_{i}\xi_{\varphi}=\frac {\partial}{\partial v}+\Omega_{i}\frac {\partial}
{\partial \overleftarrow{\varphi}}
\label{Eq:GKH}
\end{equation}
which 
 becomes null precisely over the $r=r_{i}$ hypersurfaces.
 In  fact these $r=r_{i}$  null hypersurfaces  are 
Killing horizons and their surface gravities\footnote{Our convention for the surface gravity follows  Wald«s book
ref.\cite{Wald}.}
 $k_{i}$
satisfy:
\begin{equation}
\nabla^{\mu}[g(\hat \xi_{i},\hat \xi_{i})]=-2k_{i}\hat \xi_{i}^{\mu}
\label{Eq:SGKH}
\end{equation}
where  both sides of these relations are evaluated on the $r=r_{i}$ root
of $\Delta (r)=0$. Since ${\hat g}^{rr}=0$, we get:
\begin{equation}
{\hat g}^{vr}\nabla_{r}[g(\hat \xi_{i},\hat \xi_{i})]=-2k_{i},\quad
{\hat g}^{\overleftarrow{\varphi}r}\nabla_{r}[g(\hat \xi_{i},\hat \xi_{i})]=-2k_{i}\xi^{\overleftarrow{\varphi}}
\label{Eq:SGE}
\end{equation}
The contravariant components of $g$ in the  $(v, \overleftarrow{\varphi}, r, \vartheta)$ coordinates can be evaluated
using the transformation (\ref{Eq:INC}) and we find that
\begin{equation}
{\hat g}^{vr}=\frac {I(r^{2}+a^{2})}{\rho^{2}},\quad\quad
{\hat g}^{\overleftarrow{\varphi}r}=\frac {Ia}{\rho^{2}},
\label{Eq:MC}
\end{equation}
and thus (\ref{Eq:SGE}) yield:
\begin{equation}
-2k_{i}=\frac {I(r^{2}+a^{2})}{\rho^{2}}\nabla_{r}[g(\hat \xi_{i},\hat \xi_{i})]
\label{Eq:SG1}
\end{equation}
Writing $g(\hat \xi_{i},\hat \xi_{i})=g(\xi_{t},\xi_{t})+2\Omega_{i}g(\xi_{t},\xi_{\varphi})+\Omega^{2}_{i}
g(\xi_{\varphi},\xi_{\varphi})=g(\xi_{\varphi},\xi_{\varphi})[(\Omega_{i}-a_{1})(\Omega_{i}-a_{2})]$
where $a_{1},a_{2}$ are the roots of  $g(\hat \xi_{i},\hat \xi_{i})=0$, the evaluation of the right hand side  at $r_{i}$ becomes trivial and yields:
\begin{equation}
 k_{i}= \frac {1}{2I}\frac{1}{r^{2}_{i}+a^{2}}\frac {\partial \Delta(r)}{\partial r} 
  \label{Eq:SG}
\end{equation}
where the derivative of $\Delta(r)$ is evaluated at $r=r_{i}$.
 It can be checked that the this $k_{i}$
 in the limit of $\Lambda\to 0$ reduces to the standard form of the surface gravity for the Killing horizons of 
the Kerr black hole (compare with the results  in Wald«s book \cite{Wald}).
Moreover, (\ref{Eq:SG}) shows that any Killing horizon corresponding to a double
or higher multiplicity root of $\Delta(r)=0$ is degenerate.\\\

Although above we employed ingoing coordinates to extend the  metric
in (\ref{Eq:g}) across the zeros of $\Delta(r)=0$,  an identical extension can be performed
employing outgoing coordinates. For these coordinates, we return to $(\ref{Eq:RTPND})$
 and choose the positive sign in the ''radial''
equation.
With this choice, 
 outgoing coordinates $(u, \overrightarrow{\varphi})$ are defined via 
\begin{equation}
du=dt -\frac {I(r^{2}+a^{2})}{{\Delta(r)}}dr,\quad d\overrightarrow{\varphi}=d\varphi-\frac {Ia}{{\Delta(r)}}dr
\label{Eq:ONC}
\end{equation}
Transforming again the metric $g$  in (\ref{Eq:g}) in outgoing $(u, \overrightarrow{\varphi}, r, \vartheta)$
coordinates, we get an expression analogous to (\ref{Eq:gINC}), with the only exception that 
the sign in the cross terms $(d \overrightarrow{\varphi}du)$  and 
$(d \overrightarrow{\varphi}dr)$ are reversed. \\\

We now 
use these coordinates to discuss the extendability
of causal geodesics that run into Killing horizons. 
For this let an arbitrary block $(T,g)$ specified by the condition $r\in (r_{i},r_{i+1})$,
and let us eliminate the Boyer-Lindquist coordinates $(t,r,\vartheta,\varphi)$
 in favor of 
ingoing coordinates $(v, \overleftarrow{\varphi}, r, \vartheta)$ 
where now $(v, r)$ take their values over the entire real line.
Moreover we  transform  equations $( \ref{Eq:GTM}
-\ref{Eq:GTH})$ into 
 ingoing coordinates
and after some algebra we find:
\begin{equation}
 \frac {dv}{d\lambda}=\frac {I^{2}(r^{2}+a^{2})}{\rho^{2}\Delta({r})}
 [(r^{2}+a^{2})E-al_{z}\pm \frac {\sqrt{R(r)}}{I}]+\frac {I^{2}a}{\rho^{2}{\hat {\Delta}({\vartheta})}}[l_{z}-aEsin^{2}\vartheta]
 \label{Eq:GVI}
\end{equation}
\begin{equation}
\frac {d \overleftarrow{\varphi}}{d\lambda}=
\frac {I^{2}a}{\rho^{2}\Delta({r})}[(r^{2}+a^{2})E-al_{z}\pm \frac {\sqrt{R(r)}}{I}]+ \frac {I^{2}}{\rho^{2}{\hat \Delta}({\vartheta})} [\frac {l_{z}}{sin^{2}\vartheta}-aE] 
 \label{Eq:PVI}
\end{equation}
\begin{equation}
\rho^{2} \frac {dr}{d\lambda}=\pm {R(r)}^{\frac{1}{2}},\quad \rho^{2} \frac {d\vartheta}{d\lambda}=\pm {\Theta(\vartheta)}^{\frac{1}{2}}
  \label{Eq:GRTHN}
\end{equation}

A similar computation shows that relative to the outgoing $(u, \overrightarrow{\varphi}, r, \vartheta)$ coordinates,
 $(\ref{Eq:GTM}
-\ref{Eq:GTH})$
 take the form: 
\begin{equation}
 \frac {du}{d\lambda}=\frac {I^{2}(r^{2}+a^{2})}{\rho^{2}\Delta(r)}
 [(r^{2}+a^{2})E-al_{z}\mp \frac {\sqrt{R(r)}}{I}]+\frac {I^{2}a}{\rho^{2}{\hat {\Delta}(\vartheta)}}[l_{z}-aEsin^{2}\vartheta]
 \label{Eq:OGVI}
\end{equation}
\begin{equation}
\frac {d \overrightarrow{\varphi}}{d\lambda}=
\frac {I^{2}a}{\rho^{2}\Delta(r)}[(r^{2}+a^{2})E-al_{z}\mp \frac {\sqrt{R(r)}}{I}]+ 
\frac {I^{2}}{\rho^{2}{\hat \Delta}(\vartheta)} [\frac {l_{z}}{sin^{2}\vartheta}-aE] 
 \label{Eq:OPVI}
\end{equation}
\begin{equation}
\rho^{2} \frac {dr}{d\lambda}=\pm {R(r)}^{\frac{1}{2}}, \quad 
\rho^{2} \frac {d\vartheta}{d\lambda}=\pm {\Theta(\vartheta)}^{\frac{1}{2}}
  \label{Eq:OGTHN}
\end{equation}
Let now 
$(t(\lambda), \varphi(\lambda),r(\lambda),\vartheta(\lambda))$, $\lambda \in [\lambda_{0}, \lambda_{1})$
be the coordinate representation of a causal geodesic
relative to an arbitray Carters block
and let
$(v(\lambda), \overleftarrow{\varphi}(\lambda),r(\lambda),\vartheta(\lambda))$ 
respectively $(u(\lambda), \overrightarrow{\varphi}(\lambda), r(\lambda), \vartheta(\lambda))$
with $\lambda \in [\lambda_{0}, \lambda_{1})$,
be the coordinate representation of the same geodesic relative
to ingoing-outgoing  coordinates.
We assume that this geodesic has the property that 
for $\lambda_{1}<\infty$
runs into the Killing horizon $r_{i}$
in the sense
that $\lim_{\lambda \to \lambda_{1}}r(\lambda):=r_{i}$.
As we have seen in the previous section
for this geodesic:\\

 a) either $\lim_{\lambda \to \lambda_{1}}R(r(\lambda)):=R(r_{i})>0$,
 and thus  $\lim_{\lambda \to \lambda_{1}}P(r(\lambda)):=P(r_{i})\neq0$\\
 
 b)  or  $\lim_{\lambda \to \lambda_{1}}R(r(\lambda)):=R(r_{i})=0$, $\lim_{\lambda \to \lambda_{1}}P(r(\lambda)):=P(r_{i})=0$
and $r_{i}$  is a simple zero of $\Delta(r)=0$.\\

Suppose for a particular 
choice of the constants  $(E, l^{2}, l_{z}, m),$
the geodesic 
$(t(\lambda), \varphi(\lambda),r(\lambda),\vartheta(\lambda))$, $\lambda \in [\lambda_{0}, \lambda_{1})$
satisfies condition
(a). In that event, 
equations (\ref{Eq:GVI}) and (\ref{Eq:OGVI}) can be written as:
 \begin{equation}
 \frac {dv}{d\lambda}=\frac {I^{2}(r^{2}+a^{2})P(r)}{\rho^{2}\Delta(r)}
 \left[ 1\pm \sqrt{1-\frac {\Delta(r)}{I^{2}P^{2}}}\right] +\frac {I^{2}a}{\rho^{2}{\hat \Delta}(\vartheta)}[l_{z}-aEsin^{2}\vartheta]
 \label{Eq:IFGVI}
\end{equation}
 \begin{equation}
 \frac {du}{d\lambda}=\frac {I^{2}(r^{2}+a^{2})P(r)}{\rho^{2}\Delta(r)}
  \left[ 1\mp \sqrt{1-\frac {\Delta(r)}{I^{2}P^{2}}}\right] +\frac {I^{2}a}{\rho^{2}{\hat {\Delta}(\vartheta)}}[l_{z}-aEsin^{2}\vartheta]
 \label{Eq:OFGVI}
\end{equation}
The choice of the positive sign
in  (\ref{Eq:IFGVI}) 
 implies 
that $\lim_{\lambda \to \lambda_{1}}\frac {dv}{d\lambda}$ becomes unbounded
 while the choice of
the negative sign implies that $\lim_{\lambda \to \lambda_{1}}\frac {dv}{d\lambda}$ 
has finite value\footnote{This behavior reflects the fact in one branch of of the $r_{i}$ horizon 
the  $v$-coordinate
takes its values in $(-\infty, \infty)$, while on the other branch the $v$ coordinate becomes unbounded.}.
Similar conclusions, hold for the case of equation (\ref{Eq:OFGVI}) 
with the notable difference that
in this case  $\frac {du}{d\lambda}$ becomes finite in the branch of the $r_{i}$ horizon where
$\frac {dv}{d\lambda}$ becomes unbounded and vice versa.
Moreover, since as $r\to r_{i}$, equation
(\ref{Eq:GVI}) has  the same singular 
part as that of equation (\ref{Eq:PVI}),
therefore $\frac {d \overleftarrow{\varphi}}{d\lambda}$ 
exhibits the same behavior as 
$\frac {dv}{d\lambda}$ in the limit $r\to r_{i}$
and the same conclusion 
holds  for $\frac {d \overrightarrow{\varphi}}{d\lambda}$ 
and $\frac {du}{d\lambda}$
at the same limit.\\\

Let now for a particular geodesic
$\lim_{\lambda \to \lambda_{1}}\frac {dv}{d\lambda}$ is finite.
That means $lim_{\lambda \to \lambda_{1}}(v(\lambda), \overleftarrow{\varphi}(\lambda), r(\lambda),\vartheta(\lambda))$
defines a point $q=(v_{0}, \overleftarrow{\varphi}_{0}, r_{i},  \vartheta_{0})$ on the $r_{i}$ horizon.
This point
coupled with the system (\ref{Eq:GVI}-\ref{Eq:GRTHN})
 is sufficient to extend uniquely 
 $(v(\lambda), \overleftarrow{\varphi}(\lambda),r(\lambda),\vartheta(\lambda))$ 
 as a geodesic into an adjacent block. For this extension,  we integrate (\ref{Eq:GVI}-\ref{Eq:GRTHN})
taking  $q$ as the initial point
and maintaining
the same $(E, l^{2}, l_{z}, m)$
as those that determine 
$(v(\lambda), \overleftarrow{\varphi}(\lambda),r(\lambda),\vartheta(\lambda))$ 
for 
$\lambda \in [\lambda_{0}, \lambda_{1})$.
This is a well defined operation and extends smoothly the geodesic into the adjacent block.\\

If on the other hand for a particular geodesic
turns out that 
$\lim_{\lambda \to \lambda_{1}}\frac {du}{d\lambda}$ is finite
then the same argument as above holds. 
Here the geodesic is outgoing
and defines a point $q=(u_{0}, \overrightarrow{\varphi}_{0}, r_{i},  \vartheta_{0})$ on the $r_{i}$ horizon.
This point
coupled with the system 
( \ref{Eq:OGVI}-\ref{Eq:OGTHN})
 is sufficient to extend uniquely 
 $(u(\lambda), \overleftarrow{\varphi}(\lambda),r(\lambda),\vartheta(\lambda))$ 
 as a geodesic into an adjacent block based on the same reasoning and procedure as above.\\

Finally we suppose  that 
for particular $(E, l^{2}, l_{z}, m)$ 
the geodesic 
$(t(\lambda), \varphi(\lambda),r(\lambda),\vartheta(\lambda))$, $\lambda \in [\lambda_{0}, \lambda_{1})$
obeys condition (b). Since now
 $r=r_{i}$ is a single zero of 
$\Delta(r)$ 
it follows that  $r=r_{i}$ is a single zero of $R(r)$
and thus $\lambda_{1}$ is necessary finite.
If as above,
$(v(\lambda), \overleftarrow{\varphi}(\lambda),r(\lambda),\vartheta(\lambda))$ 
respectively $(u(\lambda), \overrightarrow{\varphi}(\lambda), r(\lambda), \vartheta(\lambda))$
with $\lambda \in [\lambda_{0}, \lambda_{1})$,
are the coordinate representation of 
this geodesics in terms  of ingoing resp. outgoing coordinates, then
by appealing to (\ref{Eq:GVI}) and (\ref{Eq:OGVI}) 
and eliminating the affine parameter $\lambda$ in 
in favor of the variable $r$,
it follows that as 
$r \to r_{i}$
the right hand sides of  (\ref{Eq:GVI}) and (\ref{Eq:OGVI}) diverge simultaneously at $v\to  \infty$ and $u \to -\infty$.
This however means that the geodesic reaches a point on the bifurcation sphere
of the Killing horizon. Strictly 
this bifurcation two-sphere
it is not part of the spacetime manifold
if the latter is identified with
the chart of validity of the ingoing-outgoing coordinates.
It can be added
into the manifold 
by introducing local 
``Kruskal type coordinates`` to cover an open vicinity of the 
intersecting horizons and eventually adding the bifurcation sphere
in the same manner as for the case of a non extreme Kerr black hole
(see for instance discussion in  ref.\cite{Car3}, \cite{Neil}).\\

The  construction of
``Kruskal type coordinates`` covering the neighborhood of a non degenerate
Killing horizons
is cumbersome, 
and will not be discussed here.
Nevertheless we have checked 
that causal geodesics reaching this sphere can be continued in the same manner as for the case of 
a non extreme Kerr. This includes in particularly  the null generators of the non degenerate Killing horizons
which even though are incomplete  relative to ingoing or outgoing Finkelstein coordinates 
 their extention through the bifurcation sphere tenders them geodesicaly complete.\\

 In summary, 
causal geodesics that are reaching bifurcation spheres associated with the intersection of 
$r_{i}$-Killing horizons can  be extended smoothly as geodesics into an adjacent
blocks and incomplete null geodesics generators of these horizon are extended so that 
they become complete. Detailed discussion of the construction of ``Kruskal type coordinates`` 
for the bifurcating horizons within the Kerr-de Sitter and their properties  will be addressed elsewhere.

\section{Geodesics on the axis }

In section $IV$, we considered  an
 arbitrary block $(T,g)$ and have chosen 
the initial point $q=(t_{0}, \varphi_{0}, r_{0},\vartheta_{0})$ 
off the rotation axis and off the ring singularity
and the subsequent analysis dealt with the behavior
of such geodesics.
In this section,
we  deal with geodesics that
either pass through
or are constrained to lie along the rotation axis.
Since however, either  Boyer-Lindquist or ingoing-outgoing coordinates are 
pathological along the axis, at first we cure this pathology
by introducing 
local coordinates that cover 
the rotation axis and so that the metric is manifestly
 regular there. \\

We recall first, that in a coordinate free language, the rotation axis
is defined as the  zeros of the axial Killing
field $\xi_{\varphi}$ and it is a closed, totally geodesic submanifold that 
may consist of several disconnected components
(for a discussion and properties of totally 
geodesic submanifolds see for instance \cite{Neil}).
If we 
employ ingoing 
coordinates $(v, \overleftarrow{\varphi},r,\vartheta)$ 
then along the axis  $sin\vartheta=0$
and clearly the components of $g$ are singular
along the axis.
We restore regularity of the components by 
introducing  local coordinates $(x,y)$ via
\begin{equation}
x=sin\vartheta\cos\overleftarrow {\varphi},\quad y=sin\vartheta sin\overleftarrow {\varphi},\quad J[\frac {(x,y)}{(\vartheta,\overleftarrow\varphi)}]=cos\vartheta\sin\vartheta
  \label{Eq:Gar}
\end{equation}
where $J$ is the Jacobian determinant. 
From this transformation it follows 
\begin{equation}
d\overleftarrow{\varphi}=\frac {xdy-ydx}{x^{2}+y^{2}},\quad d\vartheta=\frac {xdx+ydy}{(x^{2}+y^{2})^{\frac {1}{2}}
(1-(x^{2}+y^{2}))^{\frac {1}{2}}}
  \label{Eq:Gif}
\end{equation}
and if $\overleftarrow{g}_{\mu\nu}$ stand for the coordinate components of $g$ relative
to the $(v, \overleftarrow{\varphi},r,\vartheta)$ chart (see (\ref{Eq:gINC})), we write:
 
 \begin{equation}
\overleftarrow{g}_{\vartheta\vartheta}d\vartheta^{2}+\overleftarrow{g}_{\varphi\varphi}d\overleftarrow\varphi^{2}=ds^{2}+[\frac {{\hat \Delta}(\vartheta)(r^{2}+a^{2})^{2}}{I^{2}\rho^{2}}sin^{2}\vartheta-
\frac {\Delta(r)a^{2}sin^{4}\vartheta}{I^{2}\rho^{2}}-\frac {\rho^{2}}{{\hat \Delta}(\vartheta)}sin^{2}\vartheta]d\overleftarrow\varphi^{2}
  \label{Eq:GAXIS}
\end{equation}
 where:
 \begin{equation}
 ds^{2}=\frac {\rho^{2}}{{\hat \Delta}(\vartheta)}[d\vartheta^{2}+sin^{2}\vartheta d\overleftarrow\varphi^{2}]
  \label{Eq:GGAXIS}
\end{equation}
Expressing  this two-metric in terms of the local coordinates $(x,y)$  we get:
\begin{equation}
ds^{2}=\frac {\rho^{2}}{{\hat \Delta}(\vartheta)}\frac {1}{1-x^{2}-y^{2}}[dx^{2}+dy^{2}-(xdy-ydx)^{2}]
  \label{Eq:1GAXIS}
\end{equation}
which is regular at $x=y=0$.
Moreover it is seen easily that
the coefficient of the term $d\overleftarrow\varphi^{2}$ in the right handside of (\ref{Eq:GAXIS})
vanishes as the axis is approached which implies that the metric $g$
in (\ref{Eq:gINC})
when expressed in terms of $(x,y)$
is regular near and on the axis $x=y=0$. Furthermore from (\ref{Eq:Gar}), it follows  that
\begin{equation}
\xi_{\varphi}=\frac {\partial}{\partial \overleftarrow{\varphi}}=x \frac {\partial}{\partial y}-y\frac {\partial}{\partial x}
\label{Eq:Kcar}
\end{equation}
which implies that relative to the new coordinates,
one component  of the axis, denoted by $A(K-\Lambda)$,
 is the set:
\begin{equation}
A(K-\Lambda)=  \left( (v, r, x=y=0), (v, r)\in (-\infty,\infty) \right)
\label{Eq:Axis}
\end{equation}

We now use the local chart $(v,r, x, y)$ to study geodesics
near and on the axis. The transformation 
(\ref{Eq:Gar}) implies
that the coordinate components of a geodesic 
 away from the axis satisfy:
\begin{equation}
\frac {dx}{d\lambda}= [\frac {x^{2}}{x^{2}+y^{2}}-x^{2}]^{\frac {1}{2}} \frac {d\vartheta}{d\lambda}-y\frac {d\overleftarrow{\varphi}}{d\lambda}
\quad\quad
\frac {dy}{d\lambda}= [\frac {y^{2}}{x^{2}+y^{2}}-y^{2}]^{\frac {1}{2}} \frac {d\vartheta}{d\lambda}+x\frac {d\overleftarrow{\varphi}}{d\lambda}
\label{Eq:CXY}
\end{equation}
\begin{equation}
 \frac {dv}{d\lambda}=\frac {I^{2}(r^{2}+a^{2})}{\rho^{2}\Delta({r})}
 [(r^{2}+a^{2})E\pm \frac {\sqrt{R(r)}}{I}]-\frac {I^{2}a^{2}E}{\rho^{2}{{\hat \Delta}(x,y)}}(x^{2}+y^{2})
 \label{Eq:GVIA}
\end{equation}

\begin{equation}
\rho^{2} \frac {dr}{d\lambda}=\pm {R(r)}^{\frac{1}{2}}  \label{Eq:GRTHNA}
\end{equation}

where $\frac {d{\vartheta}}{d\lambda}$ and $\frac {d\overleftarrow{\varphi}}{d\lambda}$
are given by (\ref{Eq:GRTHN}), (\ref{Eq:PVI})
with the understanding that 
in the right hand side of these equations $(sin\vartheta, cos\vartheta)$ should be 
eliminated in favor of the $(x,y)$-coordinates.\\

One family of solutions of 
(\ref{Eq:CXY},  \ref{Eq:GVIA},  \ref{Eq:GRTHNA})
represent geodesics  that cross  the axis.
Since $\xi_{\phi}=0$ along  the axis,
 such geodesics must have:  $l_{z}=0$
a conclusion that we already arrived
in section $II$. Moreover, we choose the other constants
 $(E,l^{2},  m^{2})$ so that $\Theta(\vartheta)$ and  $R(r)$ are
   positive on and near the axis. Positivity of
  $\Theta(\vartheta)$ on and near the axis is 
  guaranteed provided $l^{2}-m^{2}a^{2}>0$ while from
  $R(r)=I^{2}(r^{2}+a^{2})^{2}E^{2}-{\Delta(r)}(m^{2}r^{2}+l^{2})$
it is not difficult to arrange positivity of $R(r)$ for any $r_{0}$ on the axis.
Since $x\frac {d\overleftarrow{\varphi}}{d\lambda}$ 
and $y\frac {d\overleftarrow{\varphi}}{d\lambda}$ are vanishing as 
the axis is approached, we may neglect their contribution in (\ref{Eq:CXY})
and thus we get  from (\ref{Eq:CXY})
\begin{equation}
(\frac {dy(x)}{dx})^{2}\simeq\frac {y^{2}}{x^{2}}
\label{Eq:SLXY}
\end{equation}
whose solution\footnote{In the above 
and in the following equations the symbol $ \simeq$ signifies that 
only leading terms are reported, higher order terms that 
 are vanishing as the axis is approached have been neglected.}
has the form  $y(x)=\pm A^{2}x +O(x^{3})$
where $A$ is a non vanishing constant.
This solution in turn implies $x(\lambda)\simeq A_{1}\lambda$
and $y(\lambda)\simeq A_{2} \lambda$ with $A_{1}^{-1}A_{2}=A^{2}$
and with these local solutions,
(\ref{Eq:GVIA}) and (\ref{Eq:GRTHNA}) can be integrated.
Clearly the resulting solution represents a geodesic that passes through the axis.
Are these geodesics complete? Since they  
are not equatorial and thus necessary avoid the ring singularity, 
 one expects that indeed they are. 
However a proof 
of  this property it is not obvious
since the local coordinate chart employed above
covers only a vicinity of the axis.\\
Another important problem associated with  geodesics
intersecting the axis, concerns polar or polar spherical geodesics
ie geodesics that cross more than once
the polar axis. The Kerr metric admits such geodesics (see for instancce \cite{Tso}, \cite{The}), but does this property
holds for the case of Kerr de Sitter metric?
Once again one expects an affirmative answer 
 but as far as we are aware  a detailed proof is lacking.
We are hoping to return to these problems in a future work.

Geodesics that are 
restricted to lie on the axis
satisfy a simpler set
of equations.
The requirement that a  geodesic lies on the axis demand 

\begin{equation}
\frac {dx(\lambda)}{d\lambda}=\frac {dy(\lambda)}{d\lambda}=0.
\label{Eq:xdot}
\end{equation}
But
(\ref{Eq:CXY})
implies  that along the  axis
\begin{equation}
\frac {dx(\lambda)}{d\lambda}=[\frac {1}{1+(\frac{dy}{dx})_{0}^{2}}]^{\frac {1}{2}}\frac {d\vartheta}{d\lambda},\quad
\frac {dy(\lambda)}{d\lambda}=[\frac {(\frac{dy}{dx})_{0}^{2}}{1+(\frac{dy}{dx})_{0}^{2}}]^{\frac {1}{2}}\frac {d\vartheta}{d\lambda}
\label{Eq:xdott}
\end{equation}
and these eqs combined with ( \ref{Eq:TH})
implies that 
$\frac {d\vartheta(\lambda)}{d\lambda}=0$ along the axis, 
provided  $l^{2}=m^{2}a^{2}$.
This condition coupled with $l_{z}=0$ 
gives $R(r)=I^{2}[(r^{2}+a^{2})E]^{2}-\Delta(r) m^{2}(r^{2}+a^{2})$
and thus we find the following equations obeyed by $v(\lambda), r(\lambda)$
along the axis:
 
\begin{equation}
\frac {dv(\lambda)}{d\lambda}=\frac {I(r^{2}+a^{2})^{2}}{\Delta(r)}
[E\pm(E^{2}-\frac {\Delta(r)m^{2}}{I^{2}(r^{2}+a^{2})})^{\frac {1}{2}}],\quad
\frac {dr(\lambda)}{d\lambda}=\pm [I^{2}E^{2}-
\frac {\Delta(r)m^{2}} {(r^{2}+a^{2})}]^{\frac {1}{2}}
\label{Eq:GA}
\end{equation}

These equations in the limit of vanishing $\Lambda$ agree with the equations derived
by Carter in ref. \cite{Car4}.
For $m=0$, the solutions of that system can be expressed in terms of 
elementary functions. The second equation implies that a multiple of $r$ can be taken as an affine parameter
and the first equation depending upon the roots of $\Delta(r)=0$ involves only elementary integrals.
For $m\neq 0$ a great deal about the behavior of the solutions can be obtained from the second equation
which can be written in the form: 

\begin{equation}
(\frac {dr(\lambda)}{d\lambda})^{2} +V(r)=I^{2}E^{2},\quad\quad V(r)=\frac {\Delta(r)m^{2}} {(r^{2}+a^{2})}
\label{Eq:EE}
\end{equation}
which can be interpreted as describing 
the motion of a fictitious particle moving in one dimension
under the action of the potential $V(r)$. For $\Lambda >0$,
the potential exhibits a repulsive barrier  in the sense
that particles injected from one asymptotic region
will reach the other asymptotic region provided that
they have sufficient ``energy`` to overcome the height of the barrier.
In that respect Kerr-de Sitter behaves as ordinary Kerr
ie the ring like singularity for both spacetimes appears to be repulsive. 
However it is worth to mention that for the  Kerr-de Sitter
$\Delta(r)$ is a quartic polynomial, and here there is the possibility that
particles states can get  trapped between the local maxima of $V(r)$
and this effect is absent for the case of ordinary Kerr.
We shall discuss these implications  in a future work.

\section{Discussion}

In this work, we presented an analysis of
timelike and null geodesics on an arbitrary Kerr-de Sitter spacetime
and a key ingredient for this 
analysis was
 the IVP  defined in (\ref{Eq:IVPRT1}).
Even
 though
 this IVP is defined 
relative to a Carters block equipped with
Boyer-Lindquist coordinates, nevertheless
we have been able to show that these geodesics
can be continued in an unambiguous manner
through the  Killing horizons.
  Our method, in the limit $\Lambda \to 0$
 applies to the description of geodesics for a Kerr 
 background and in a sense the method complements the original treatment of Carter
 \cite{Car3}.\\
 The results of the present paper
offer only a first glimpse
on the behavior of causal geodesics
on a Kerr-de Sitter spacetime.
A detailed analysis of the behavior of these geodesics 
is a problem of immense complexity.
Even the comparatively simpler problem of the behavior
geodesics on a Kerr background has 
been adequately understood only after
intense scrutiny  for number of years by many people
(for an update on the current state of the development 
on the behavior of geodesics on Kerr consult  the monograph by Neil 
\cite{Neil}).
In contrast, the behavior of geodesics on a Kerr-de Sitter is largely
unexplored territory.
The present work offers a few tools for further exploration.
Based on the present analysis, we may 
address issues  
 like the global behavior
of these geodesics.
Where are the endpoints of these causal geodesics?
An analysis of this problem requires an understanding the global structure of 
Kerr-de Sitter and the nature of their maximal analytical
extension. Judging from
the
complicated nature 
of the Carter-Penrose conformal diagrams of 
the rotation axis (\cite{GibHaw1},\cite{Mat1})
one expects
a complicated global structure
and as far as we are aware that problems is still open.\\

Furthermore the present analysis 
offers tools for a systematic investigation 
of specific families of geodesics 
such as equatorial geodesics,  spherical causal geodesics or polar orbits
We hope to discuss their properties 
in a future work.\\

The behavior of causal geodesics on a Kerr-de Sitter
offers insights regarding  cosmic censorship
 within the cosmological domain.
 This connection arises 
by noting that the family of Kerr-de Sitter includes
spacetimes where observers find themselves enclosed within a pair of
cosmological horizons and exposed to the influence of the ring-like curvature singularity.
Is this ring like singularity visible by static observers? 
An investigation of that problem is of interest, although
it is a problem of considerable mathematical difficulty.
The difficulty arises from the fact that the IVP
defined
in (\ref{Eq:IVPRT1}) becomes singular when the initial point is chosen on the ring singularity
and thus the well known theorems regarding the existence and uniqueness of solutions of this IVP 
break down. Here the situation is analogous
to the problem of predicting the nature of causal geodesics emanating from the shell focusing singularity in the
 Bondi-Tolman collapse (see for instance ref.\cite{OS} and references therein).
While for the case of a shell focusing singularity, null geodesics with non vanishing angular momenta 
 are emerging from the singularity, it would be worth while to investigate whether geodesics with non vanishing 
  $Q$ emerge from the ring singularity
  of a Kerr-de Sitter (or a Kerr) spacetime. \\ 

\acknowledgments
We thank the members of the relativity group at the Universidad Michoacana for stimulating discussions.
Special thanks are due to O. Sarbach for his interest in this work and for his constructive criticism.
T.Z would like to thank Kayll Lake for many discussions on Kerr and Kerr de Sitter spacetimes and 
for  his help to evaluate curvature invariants using GRTensorII.
The research of T.Z was supported in part by CONACyT Grant  No. 271904 and by a CIC Grant from the
Universidad Michoacana, Mexico, J.F.S thanks CONACyT for a post-graduate Fellowship.

\section{Appendix I}
In this Appendix, we address the issue of the existence and uniqueness
of solutions of the IVP problem
 defined in (\ref{Eq:NIVP})
or the closely related IVP defined in 
 (\ref{Eq:NNIVP}).
 We recall that we run into these IVPs during the process of continuation of solutions through
 the boundary $\partial D$ of a given domain $D$.
 We treat first the case of
 (\ref{Eq:NIVP})
 which for the readers convienience we rewrite this IVP:
\begin{equation}
\rho^{2} \frac {dr}{d\lambda}= {R(r)}^{\frac{1}{2}},\quad \rho^{2} \frac {d\vartheta}{d\lambda}= -{\Theta(\vartheta)}^{\frac{1}{2}},\quad r( \lambda_{1})=\hat r_{0},\quad \vartheta(\lambda_{1})=\vartheta_{2}
 \label{Eq:NIVPd}
\end{equation}

 As we stressed in the main text 
  since $\Theta(\vartheta_{2})=0$ which implies  that
$ {\Theta(\vartheta)}^{\frac{1}{2}}$ 
fails to be continuously differentiable  
in an open vicinity of the point:
$ \vartheta(\lambda_{1})=\vartheta_{2}$
we cannot  apply standard existence and uniqueness theorems for IVPs 
to conclude the existence of a unique solution.
Here we show that as long as
${\Theta(\vartheta)}$ has a simple zero at $\vartheta_{2}$
and $\vartheta(\lambda)=0$ has a turning point at
$\vartheta_{2}$, exists an $[\lambda_{1}, \lambda_{2})$ 
and unique functions 
$\vartheta(\lambda), r(\lambda)$
satisfying (\ref{Eq:NIVPd}).
\\

In order to establish this property, at first based on $R(\hat r_{0})\neq 0$, we can in an open vicinity of $\hat r_{0}$
eliminate the parameter $\lambda$
in favor of $r$ and thus consider the equivalent system:

\begin{equation}
 \frac {d\vartheta(r)}{dr}=-\frac {[{\Theta(\vartheta(r)}]^{\frac{1}{2}}}{{R(r)}^{\frac{1}{2}}},\quad \vartheta(\hat r_{0})=\vartheta_{2},\quad r\geq \hat r_{0}
 \label{Eq:NIVPD}
\end{equation}
Since $\Theta(\vartheta)$ has a simple zero at $\vartheta_{2}$,
we write:
\begin{equation}
\Theta(\vartheta(r))=(\vartheta_{2}-\vartheta(r))G(\vartheta(r)),\quad G(\vartheta_{2})>0,\quad r\geq \hat r_{0}
 \label{Eq:EXPA}
\end{equation}
for some smooth positive function $G(\vartheta)$ for $\vartheta$ on some interval $[\theta_{2}, \hat \theta)$.

Using this expansion in  (\ref{Eq:NIVPD}), we find that the function: $A(r)=(\vartheta_{2}-\vartheta(r))^{\frac{1}{2}}, r \geq \hat r_{0}$
satisfies:
\begin{equation}
 \frac {dA(r)}{dr}=\frac{1}{2} \frac {{G(r)}^{\frac{1}{2}}}{{R(r)}^{\frac{1}{2}}},\quad A(\hat r_{0})=0,\quad r\geq \hat r_{0}
 \label{Eq:NIVPB}
\end{equation}
Since the right hand side of this equation is smooth and strictly positive
in the open vicinity of $\hat r_{0}$,
exists an interval $[\hat r_{0}, r_{1})$
and a unique, smooth and positive  for $r >\hat r_{0}$, function $A(r)$ that satisfies on 
$[\hat r_{0}, r_{1})$
the above IVP. This in turn implies that
the unique  function $\vartheta(r)=\vartheta_{2}-A^{2}(r)$ 
satisfies 
 $\vartheta(\hat r_{0})=\vartheta_{2}$ 
and on $[\hat r_{0}, r_{1})$ the IVP described by (\ref{Eq:NIVPD}).

We can easily now show that these considerations lead to 
the establishing a unique solution of 
 (\ref{Eq:NIVPd}). Indeed from the knowledge of $\vartheta(r)$ and $R(r)$
 we define:
 \begin{equation}
 \lambda(r)=\int_{\hat r_{0}} ^{r}\frac {s^{2}+a^{2}cos^{2}\vartheta(s)}{R(s)^{\frac {1}{2}}}ds +con, \quad r\in [\hat r_{0}, r_{1})
  \label{Eq:AF}
\end{equation}
 and adjust the constant 
 so that the inverse function $r(\lambda), \lambda \in [\lambda_{1},\lambda_{2}) $ satisfies $r(\lambda_{1})=\hat r_{0}$.
 Next, we introduce\footnote{Strictly we should write
 $\hat \vartheta(\lambda)=\vartheta (r(\lambda))$, but 
 with the usual ambush of notation we drop the hat.} $\vartheta(\lambda)=\vartheta (r(\lambda)), \lambda \in [\lambda_{1},\lambda_{2}) $ 
 and it is clear by constriction the pair 
 of functions $\vartheta(\lambda), r(\lambda), \lambda \in [\lambda_{1},\lambda_{2}) $  
 defined above is a smooth solution of (\ref{Eq:NIVPd}).\\
 
 A similar type of analysis, holds for the case of the IVP defined
 in (\ref{Eq:NNIVP}). As long as $R(r)$ has a simple root at $\hat r_{2}$,
 then the above analysis can be repeated with $\Theta(\vartheta)$ been replaced
 by $R(r)$ and eventually conlclude the existence of a unique solution
 for the IVP defined in (\ref{Eq:NNIVP}).

\section{Appendix II}

In this second appendix, we 
consider
 the IVP posed in (\ref{Eq:NIVP})
 but currently
 assuming that 
 the constants
$(l^{2},E,  l_{z}=0, m \geq 0)$ have been chosen so that 
$\Theta(\vartheta)>0$ on the entire interval $[0,\pi]$.
Positivity of $\Theta(\vartheta)$ on $[0,\pi]$ can be imposed since
under the assumption $l_{z}=0$, it follows from 
(\ref{Eq:THN}) that 
\begin{equation}
\Theta(\vartheta)= ax^{2}+bx+Q,\quad x=cos^{2}\vartheta,\quad a=-\frac {\Lambda a^{2}}{3}m^{2}a^{2},\quad
b=a^{2}(I^{2}E^{2}-m^{2})+\frac {\Lambda a^{2}}{3}(Q+I^{2}a^{2}E^{2}).
\label{Eq:THN1}
\end{equation}
It can be verified, that for $Q>0$
and say $I^{2}E^{2}>m^{2}$, we can arrange matters so that
the positive root $x_{+}$ satisfies  $x_{+}>1$
and thus 
the function $\Theta$ can be positive on the entire interval $[0,\pi]$.
Our goal in this appendix, is to offer a few comments on
the solutions of the IVP:
 \begin{equation}
\rho^{2} \frac {dr}{d\lambda}= \pm{R(r)}^{\frac{1}{2}},\quad \rho^{2} \frac {d\vartheta}{d\lambda}= \pm{\Theta(\vartheta)}^{\frac{1}{2}},\quad r( \lambda_{1})=\hat r_{0},\quad \vartheta(\lambda_{1})=\vartheta_{2}\quad (r,\vartheta) \in (\hat r_{1}, \hat r_{2})\times[0,\pi]. 
 \label{Eq:NIVPA}
\end{equation}
under the assumption that $\Theta(\vartheta)$
is nowhere vanishing on $[0,\pi]$.
For simplicity, we assume that 
 $r$ is restricted to a block subject to $\Delta(r)>0$
 and $R(r)=0$ admits  two positive, simple roots $\hat r_{1}< \hat r_{2}$
 with $R(r)>0$ on $(\hat r_{1}, \hat r_{2})$.
For this case, 
we eliminate the affine parameter $\lambda$ 
 in (\ref{Eq:NIVPA}) in favor of $\vartheta$,
and we find that  $r(\vartheta)$ satisfies: 

 \begin{equation}
\frac {dr(\vartheta)}{d\vartheta}=\pm\frac {[R(r(\vartheta)]^{\frac{1}{2}}}{\Theta({\vartheta})^{\frac {1}{2}}},\quad r(\theta_{0})=
\hat a,
\quad (\hat a,\vartheta_{0}) \in (\hat r_{1}, \hat r_{2})\times [0,\pi],\quad \vartheta \in [0,\pi]. 
 \label{Eq:NIVPB}
\end{equation}
Since the right hand side of this equation is bounded and continuously differentiable on $(0,\pi)$, any solutions of this  IVP can be extended up to the boundaries $r=\hat r_{1}$ ,  $r=\hat r_{2}$
or $\vartheta=0$ , $\vartheta=\pi$. If for instance, we consider the positive sign  in 
(\ref{Eq:NIVPA}), and thus the positive sign in (\ref{Eq:NIVPB}), then if the solution 
 reaches the $r=\hat r_{2}$ component of the boundary, 
 in the sense $r({\hat \vartheta_{2}})=\hat r_{2}$ for an ${\hat \vartheta_{2}} \in (0,\pi)$,
 then from the results of Appendix $I$,
 we continue the $r(\vartheta)$ for  $\vartheta\geq{\hat \vartheta_{2}}$ 
 as the unique solution of the IVP:
  \begin{equation}
\frac {dr(\vartheta)}{d\vartheta}=-\frac {[R(r(\vartheta))]^{\frac{1}{2}}}{\Theta({\vartheta})^{\frac {1}{2}}},\quad 
r({\hat \vartheta_{2}})=\hat r_{2}\quad
\vartheta\in [\hat \vartheta_{2},\pi). 
 \label{Eq:NIVPE}
\end{equation}
 From the results of Appendix $I$,
 this is a well defined IVP whose solution can be extended until it 
 reaches either of the components $r=\hat r_{1}$ or 
$\vartheta=\pi$ of the boundary. In the 
former case the solution can be 
continued in the familiar manner while 
for the latter case the solution reaches
a point on the rotation axis
and here the results of section $VI$ are applicable.\\

It is worth however to point out one family of solutions of 
the IVP (\ref{Eq:NIVPA}), or the equivalent (\ref{Eq:NIVPB}).
Suppose that there exist initial data so that a solution $r(\vartheta)$ of
(\ref{Eq:NIVPB}) satisfy:\\

a) $r(\vartheta)$ exist for all $\vartheta \in [0,\pi]$\\

b) $r(\vartheta)$ has a turning point at an $\hat \vartheta \in (0, \pi)$\\

then this solution is part of a polar orbit that
crosses the north axis at $r_{N}=r(0)$, the south axis
at $r_{S}=r(\pi)$ and it is bounded by 
$r_{B}=r(\hat \vartheta)$.
Although this argument implies that polar orbits should exist, needless to say more work is needed to
reveal their properties. We hope to return to these issue in a future work.

\end{document}